\documentclass[twocolumn,aps,amssymb,epsfig,paper]{revtex4}
\usepackage{epsfig}
\usepackage{wasysym}

\def\figswidth{58mm}
\def\figsskip{-1.5mm}
\def\question{{(Unk.)}}
\def\nsystems{44\,\,}
\def\vnegspace{\vspace{-3mm}}

\begin{document}
\textheight 226.5mm
\title{\Large {Hafnium binary alloys from experiments and first principles}}
\author{
Ohad Levy$^{1,2}$, Gus L. W. Hart$^{3}$, and Stefano Curtarolo$^{1,\star}$}
\affiliation{
$^1$Department of Mechanical Engineering and Materials Science and
Department of Physics, Duke University Durham, NC 27708 \\
$^2$Department of Physics, NRCN, P.O.Box 9001, Beer-Sheva, Israel\\
$^3$Department of Physics and Astronomy, Brigham Young University, Provo UT 84602\\
$^\star${corresponding author, e-mail: stefano@duke.edu}
}
\date{\today}
\begin{abstract}
  Despite the increasing importance of hafnium in numerous technological applications, experimental
  and computational data on its binary alloys is sparse.  In particular, data is scant on those
  binary systems believed to be phase separating.  We performed a comprehensive study of \nsystems
  hafnium binary systems with alkali metals, alkaline earths, transition metals and metals, using
  high-throughput first principles calculations. These computations predict novel unsuspected
  compounds in six binary systems previously believed to be phase separating. They also predict a few
  unreported compounds in additional systems and indicate that some reported compounds may actually
  be unstable at low temperatures.  We report the results for the following systems: AgHf, AlHf, AuHf,
  BaHf$^\star$, BeHf, BiHf, CaHf$^\star$, CdHf, CoHf, CrHf, CuHf, FeHf, GaHf, HfHg, HfIn, HfIr, HfK$^\star$,
  HfLa$^\star$, HfLi$^\star$, HfMg, HfMn, HfMo, HfNa$^\star$, HfNb$^\star$, HfNi, HfOs, HfPb, HfPd,
  HfPt, HfRe, HfRh, HfRu, HfSc, HfSn, HfSr$^\star$, HfTa$^\star$, HfTc, HfTi, HfTl, HfV$^\star$,
  HfW, HfY$^\star$, HfZn, and HfZr\\
  ($^\star$ = systems in which the {\it ab initio} method predicts that no compounds are stable).
\end{abstract}
\maketitle

High-throughput (HT) calculations of material properties based on density functional theory (DFT)
have been developed in recent years for theoretically guided material discovery and improvement
\cite{Johann02,Stucke03,curtarolo:prl_2003_datamining,monster,Fischer06,Lewis2007355,ozolins:135501,Ortiz09}.
These calculations give insights into trends in alloy properties and indicate possible existence of
hitherto unobserved compounds. In this paper we apply the HT approach to a comprehensive screening
of hafnium intermetallic binary alloys. This choice is motivated by the wide array of technological
applications of hafnium alloys in contrast with their scant theoretical discussion in the
literature.

Hafnium is primarily used in the control and safety mechanisms of nuclear reactors, because of its
high cross-section for neutron absorption and its high corrosion resistance \cite{ASMHandbook}.
Hafnium cladding of nuclear fuel rods is expected to be an important
element in the design of future advanced reactors \cite{Wallenius08}.
Hafnium is used extensively as an alloying element in
nickel-, niobium-, and tantalum-based superalloys, which are designed to withstand high temperatures
and pressures. It is an important addition to some titanium, tungsten and molybdenum alloys, where
it forms second-phase dispersions (with carbon) that improve material strength under extreme
conditions \cite{ASMHandbook,Nielsen_Hf}.
Hafnium alloys are also used in medical implants and devices, due to their biocompatibility and
corrosion resistance (see for example \cite{patent_5954724}). Nickel-titanium-hafnium alloys
exhibit shape memory behavior with high martensitic transformation temperatures and good mechanical
properties \cite{Meng09}.  Hafnium is added to aluminum-magnesium-scandium alloys, widely used in
aerospace applications, to increase their strength following high temperature thermomechanical
processing \cite{Patent_012868}.  Some intermetallic compounds of Hf and the transition metals Fe,
Co, Pd and Pt have been investigated as hydrogen-storage materials because of their capability to
form hydrides with high hydrogen to metal ratios at room temperature \cite{Baudry92}.  

Hafnium oxide based compounds have recently found wide application replacing silicon oxide as
high-$k$ dielectrics in the production of integrated circuits \cite{Callegari04}. This has motivated
a few first principles studies of the dielectric properties of hafnium silicates (see for example
\cite{Pignedoli07}). Perovskite alkaline metal hafnates (e.g. CaHfO$_3$, BaHfO$_3$ and SrHfO$_3$)
have been investigated for various optical and electronic applications as well as substrates for
perovskite superconductor films, due to their high stability in the fabrication process of such
films and their small crystallographic mismatch (see e.g. \cite{Zhang94}). Although zirconium has
been tried as a cheaper substitute for hafnium, in most of these applications only hafnium
produces the desired properties \cite{ASMHandbook}.

Despite this wealth of existing and potential applications, computational studies of hafnium
compounds are few.  In most of these studies, just a few specific structures have been investigated
\cite{Ghosh02,Kong05,XQChen05,JBC06,CZhang08,Ormeci96,XQChen04,Lalic99,JBC05,Cekic04,Hao07,Shein08,XZhang08,Lalic98,Oh08,XQChen06}.
In only three cases, Al-Hf, Cu-Hf and Hf-Nb \cite{Ghosh02,Ghosh05,Ghosh08,Ghosh07}, have a
large number of
structures been examined.  In the Al-Hf system, the energies of 22 intermetallic structures have
been calculated, four of which are ground states and constitute the convex
hull of the system \cite{Ghosh05,Ghosh08} (The convex hull connects all of the ground states on
a concentration vs. enthalpy diagram. The enthalpies of all other structures lie above the tie lines
between the ground states. Thermodynamically, the convex hull represents the Gibbs free energy of the alloy at
zero temperature \cite{monster}). In the Cu-Hf system, energies of 28 structures have been
calculated and four identified as ground states \cite{Ghosh07}. In both cases, differences have been found
between the phase diagrams
based on calorimetric measurements and the calculated ground states. In contrast,
positive formation enthalpies were calculated for 13 Hf-Nb intermetallic structures \cite{Ghosh02},
in agreement with experimental observations that the system is phase-separating.

In several systems, the stability of a few
competing structures have been examined but the complete convex hull of the intermetallic system has
not been addressed. These include the ground state properties of five different structures of
Hf$_3$Mo, HfMo and HfMo$_3$ \cite{Kong05} and the three Laves phases of HfCr$_2$ \cite{XQChen05},
HfFe$_{2}$ \cite{JBC06,CZhang08}, HfV$_2$\cite{Ormeci96,CZhang08} and HfMn$_2$ \cite{XQChen04}.
Single specific structures of Hf$_2$Fe \cite{Lalic99,JBC05}, Hf$_2$Co \cite{Cekic04}, HfB$_2$
\cite{Hao07,Shein08,XZhang08}, Hf$_2$Ni \cite{Lalic98}, HfTe$_5$ \cite{Oh08}, HfRu, HfRh, HfIr and
HfPt \cite{XQChen06} have also been studied.
Some of these compounds were studied because they are used in hyperfine field
measurements by the time dependent perturbed angular correlation (TDPAC) technique, for which
$^{183}$Hf is an excellent probe, that can be easily compared to first principles calculations of the
local electric field gradient in the vicinity of the probe.

In this paper we report a comprehensive computational study of the low temperature stability of
\nsystems hafnium binary systems; hafnium combined with alkali metals, the alkaline earths,
transition metals and metals.  We apply the {\it ab initio} formation energy criterion which has
been shown to be reliable for binary systems
\cite{curtarolo:prl_2003_datamining,monster,morgan:meas_2005_ht}. 
The calculations were performed
using the high-throughput framework {\small AFLOW} \cite{monster,aflow}, employing {\it ab initio}
calculations of the energies with the VASP software \cite{kresse_vasp}. We used projector augmented
waves (PAW) pseudopotentials \cite{PAW} and the exchange-correlation functionals parameterized by
Perdew, Burke and Ernzerhof \cite{PBE} for the generalized gradient approximation (GGA). The
energies were calculated at zero temperature and pressure, with spin polarization, and without
zero-point motion or lattice vibrations.  All crystal structures were fully relaxed (cell volume and
shape and the basis atom coordinates inside the cell).  Numerical convergence to about 1meV/atom was
ensured by a high energy cutoff (40\% higher than the highest energy cutoff for the
pseudo-potentials of the components) and dense 6000 {\bf k}-point Monkhorst-Pack meshes. For each
system, we calculated the energies of all the ground state structures reported in Refs.
\cite{Massalski} and \cite{Pauling} and about 200 additional crystal structures.
In addition to the 176 structures
described in \cite{monster} these included the prototypes A5, A6, A7, A11, B20, TlI, ThIn, LiB-MS1/2
\cite{kolmogorov:binary_LiB_2006,kolmogorov:binary_borides_LiB_2006}, Au$_4$Zr, Ca$_7$Ge,
NbNi$_8$(Pt$_8$Ti), Ga$_2$Hf, W$_5$Si$_3$, V$_4$Zn$_5$, Ni$_7$Zr$_2$, C36 and the complete set of
hcp-superstructures \cite{gus_enum} with up to four atoms per cell.

\begin{figure}[h]
  \includegraphics[width=\figswidth]{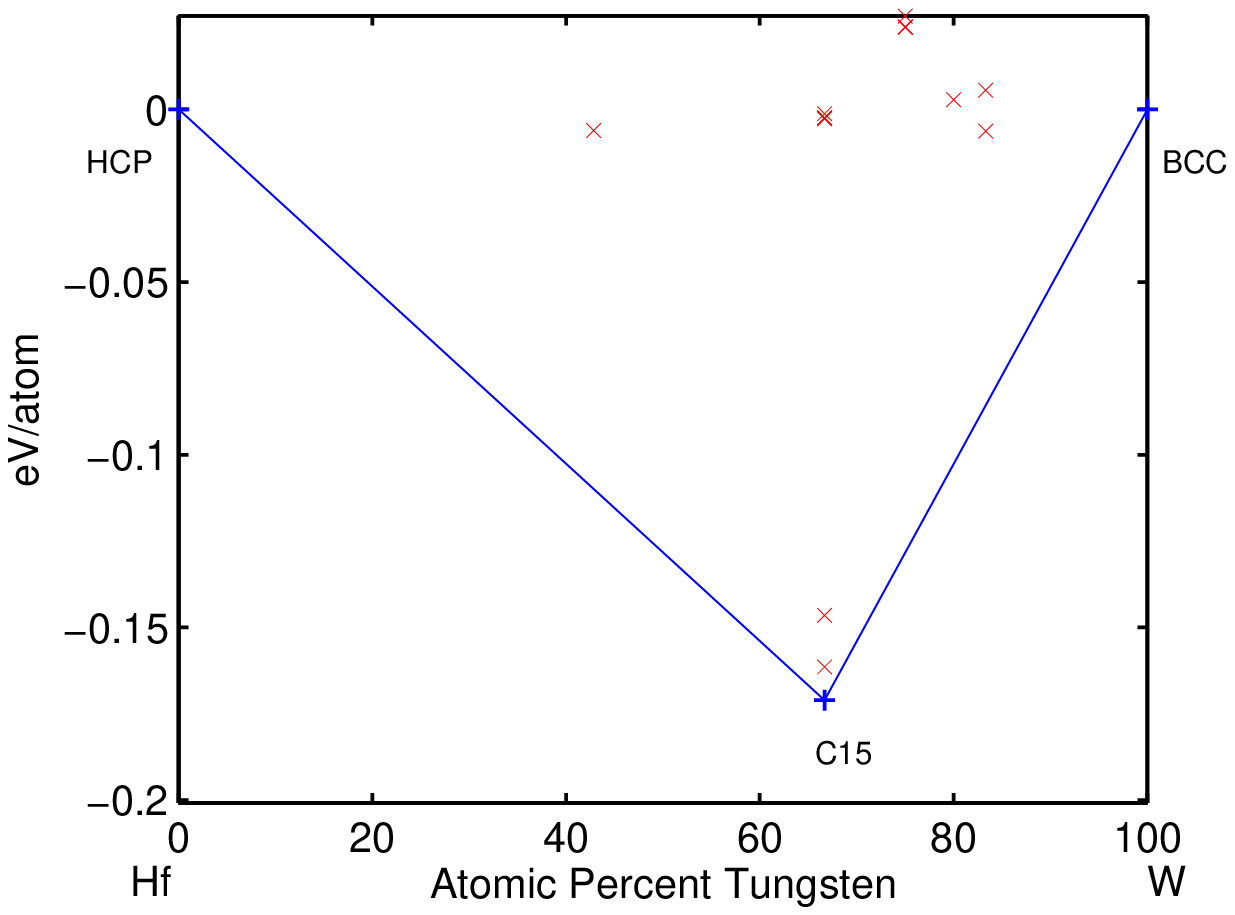}\hspace{\figsskip}
  \includegraphics[width=\figswidth]{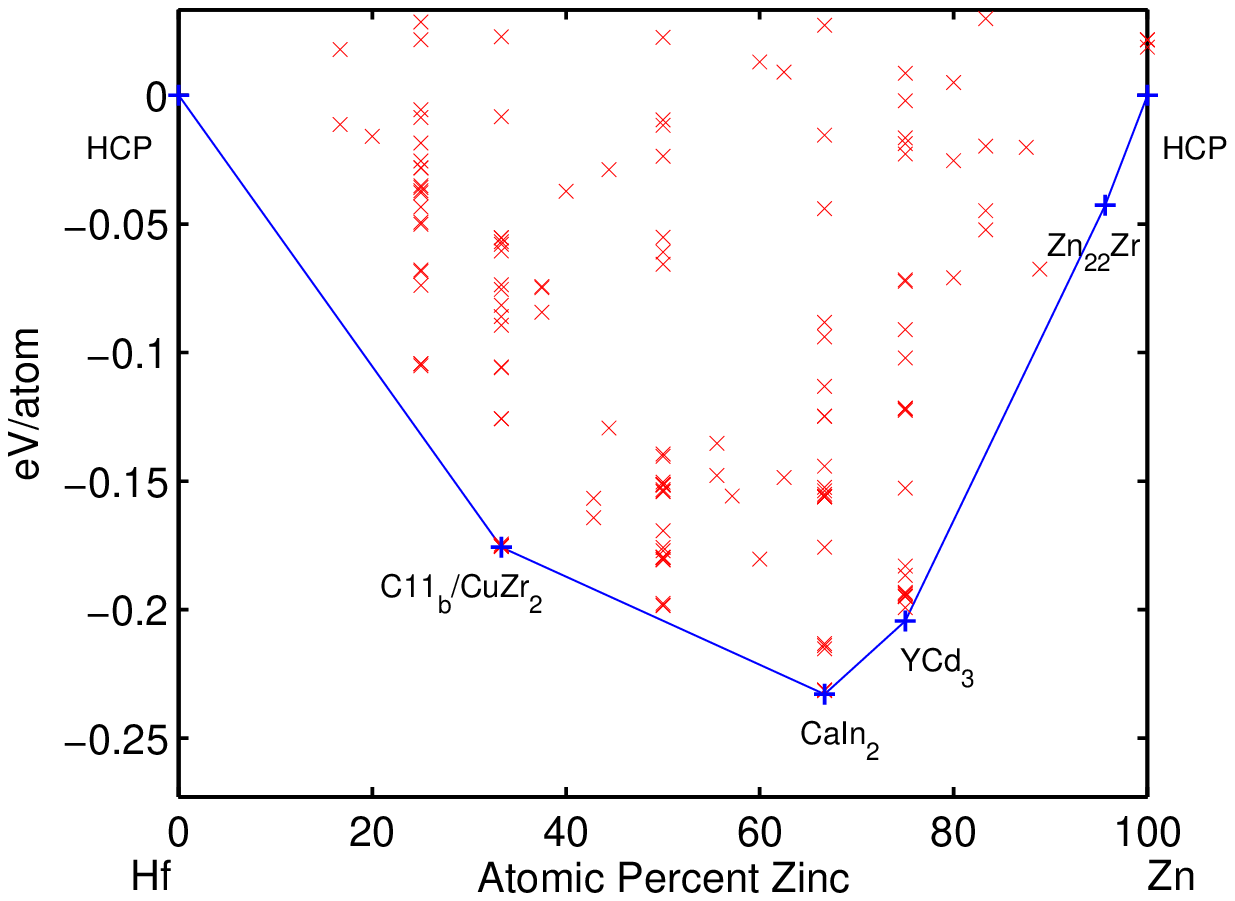}\hspace{\figsskip}
  \includegraphics[width=\figswidth]{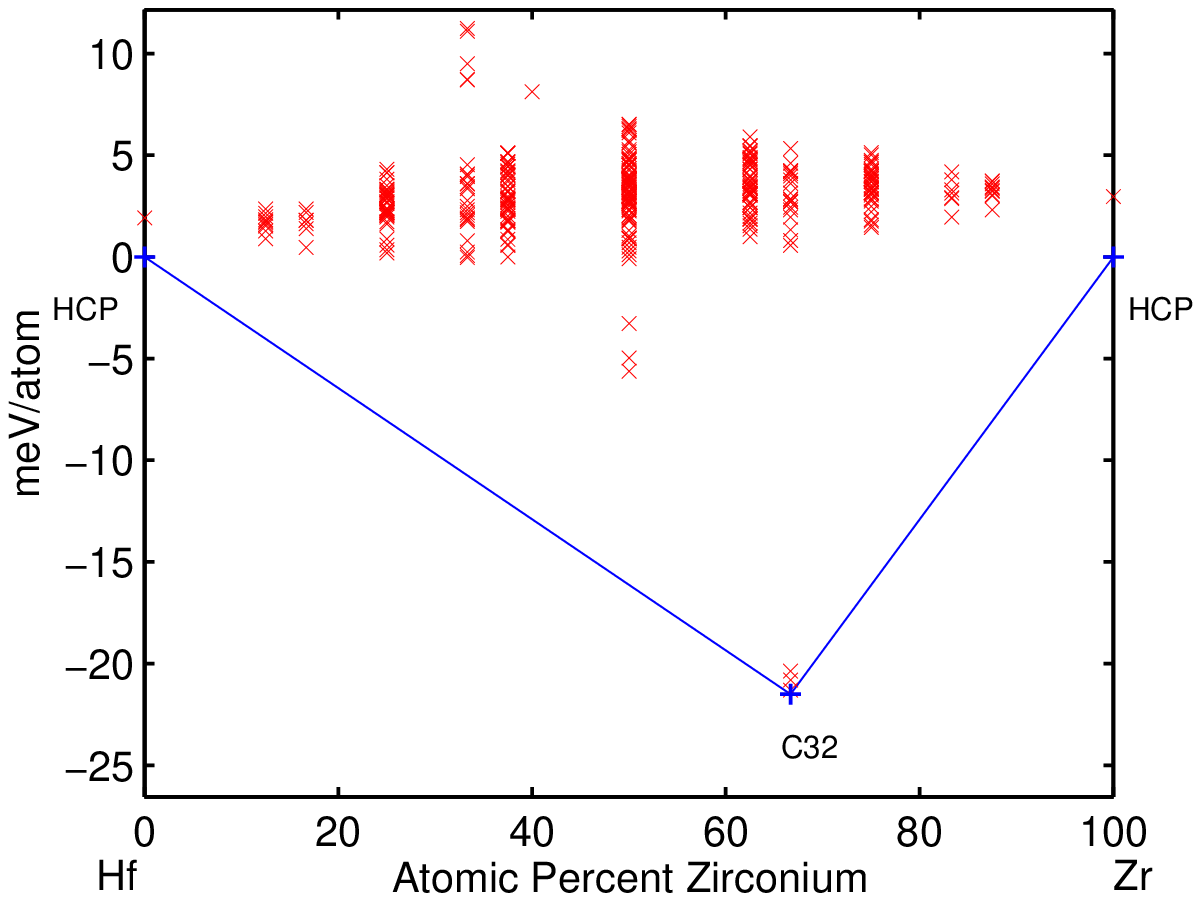}\hspace{\figsskip}
  \vnegspace
  \caption{{\small Formation enthalpies of
      Hf-W, Hf-Zn, and Hf-Zr 
      alloys.}}
  \vnegspace
  \label{fig_HfW}\label{fig_HfZn}\label{fig_HfZr}
\end{figure}

Some of our results deviate from experimental data \cite{Pauling,Massalski}. Table \ref{table1}
is a summary of the alloy systems addressed in this study.
On the first column, the alloying metals are ordered according to their
Mendeleev number (or Pettifor's chemical scale) \cite{pettifor:1984,pettifor:1986}. The reported
experimental compounds (or lack thereof) are shown in the second column and the
calculated ones in the third column. The compounds
are presented with their structure {\it Strukturbericht} designation or prototype in parentheses
(unspecified or unknown structures are denoted as Unk.). Some of the predicted phases
(marked by an asterisk)
have structures for which no prototype is known and no {\it Strukturbericht} designation have
been given. These new prototypes are described in Table
\ref{protos}. Appearance of two structures in the third column indicates
their degeneracy. The fourth column gives the calculated energies of the structures identified on
the convex hull of each binary system (3rd column). Energies of reported structures (2nd column)
that are found to lie above the convex hull (missing in the 3rd column) are indicated in
parentheses.  In cases where the reported and calculated structures of a compound are different, or
two non-degenerate structures are reported, the energy difference between them is indicated in
square parentheses. The calculated convex hulls of all the compound forming systems are shown in
Figs. (\ref{fig_AgHf})-(\ref{fig_HfZr}).

In Table \ref{table1} there are 16 systems that are reported in the literature as phase separating,
i.e. having no compounds. 13 of these systems are grouped together at the top of the table.
It is not surprising to see this same behavior in a block of the table, because the systems
are listed by Pettifor's Mendeleev number \cite{pettifor:1986}. The other three systems reported to be phase-separating,
Hf-Mg, Hf-Tl and Hf-Pb, are scattered in the lower rows of the table. In these three cases,
our calculations show that they are in fact compound forming, essentially complementing the
general trend implied by the Pettifor chemical scale. 
Stable structures are also found in the binary systems of Hf with Ti
and Zr (sharing the IVB column of the periodic table) and Sc. In five of the six systems
predicted to be compound forming (excluding Hf-Pb)
the pure elements have a hcp crystal structure and only disordered hcp solid solutions have been
reported for their binary alloys with hafnium over the entire range of concentrations. The
calculations indicate, however, that in four of them stable compounds that are \emph{not}
hcp-based superstructures should exist at low temperatures. Only in the
Hf-Sc system, the new
compounds are predicted to have a hcp based structure. Three of these metals have very high melting
temperatures and the available experimental data on their alloys is also limited to high
temperatures (above 700$^\circ$C, 800$^\circ$C, 1000$^\circ$C for Ti, Zr and Sc, respectively
\cite{Massalski,Pauling}. No data is available for Hf-Mg, Hf-Tl and Hf-Pb alloys.

\newpage

  
  \begin{table}[thw]
    \small
    \caption{
      \scriptsize
      Compounds observed in experiments
      or predicted
      by {\it ab initio} calculations
      in metallic binary alloys of Hf. Structure
      {\it Strukturbericht} designation or prototype are in parentheses (Unk.\ denotes unknown structures).
      New prototypes are marked by $\star$ and described in Table \ref{protos}. 
      More than one structure may have been reported (2nd column) or found with degenerate energies (3rd column). 
      $\Delta H$ are the formation enthalpies of the compounds in the present study; parentheses denote
      reported structures (2nd column) that lie above the convex hull (missing in the 3rd column).
      Square parentheses denote energy differences between 
      experimentally observed and calculated structures or two reported non-degenerate structures. 
    } \label{table1}
    \scriptsize
    \begin{tabular}{c|cc|c}
    \hline \hline
& \multicolumn{2}{c|}{Compounds} & $\Delta H$ \\
& Experiments \cite{Pauling,Massalski} & Calculations & meV/at\\ \hline
 K,Na  &  -  & - &  \\ \hline
 Li,Ba &  -  & - &  \\ \hline
 Sr,Ca,Y  &  -  & - &  \\ \hline
 Sc &  -  & Hf$_5$Sc$^\star$ & -10 \\ 
    &     & Hf$_3$Sc$^\star$ & -10 \\ \hline
 La &  -  & - &  \\ \hline
 Zr &  -  & HfZr$_2$(C32) & -22 \\ \hline
 Ti &  -  & HfTi$_2$(C32) & -11 \\ \hline
 Nb,Ta &  -  & - &  \\ \hline
 V  & HfV$_2$(C15) & - & (31) \\ \hline
 Mo & HfMo$_2$(C15) & HfMo$_2$(C15) & -170 \\ \hline
 W  & HfW$_2$(C15) & HfW$_2$(C15) & -171 \\ \hline
 Cr & HfCr$_2$(C15, C36) & HfCr$_2$(C15)  & -120[+6] \\ \hline
 Tc &    & Hf$_3$Tc(Mo$_3$Ti$^\star$ \cite{monster})& -269 \\
    &    & Hf$_2$Tc(C49) & -357\\ 
    & HfTc(B2) & HfTc(B2) & -482 \\
    & HfTc$_2$(C14) & HfTc$_2$(C14)& -362\\  \hline
 Re &    & Hf$_3$Re(Mo$_3$Ti$^\star$ \cite{monster}) & -200 \\ 
    & HfRe \question\, \cite{Pauling} &  & \\
    & Hf$_{21}$Re$_{25}$(Re$_{25}$Zr$_{21}$) \cite{Massalski} & Hf$_{21}$Re$_{25}$(Re$_{25}$Zr$_{21}$) & -407 \\ 
    & HfRe$_2$(C14) & HfRe$_2$(C14) & -394 \\ 
    & Hf$_5$Re$_{24}$(Ti$_5$Re$_{24}$) & Hf$_5$Re$_{24}$(Ti$_5$Re$_{24}$) & -252\\ \hline
 Mn & HfMn$_2$(C14, C36) & HfMn$_2$(C14) & -268[+4] \\
    & Hf$_2$Mn(NiTi$_2$) &   &  (-109) \\ \hline
 Fe &    &  Fe$_5$Hf(C15$_b$) & -178 \\
    & Fe$_2$Hf(C14, C15) & Fe$_2$Hf(C15) & -354[+14] \\
    &    & FeHf(B2) & -331 \\
    & FeHf$_2$(NiTi$_2$) &  & (-189) \\ \hline
 Os & Hf$_{54}$Os$_{17}$(Hf$_{54}$Os$_{17}$) &  & (-319) \\
    & Hf$_2$Os(NiTi$_2$) &  & (-429) \\
    & HfOs(B2) & HfOs(B2)& -707\\  
    & HfOs$_2$(C14) &  & (-402) \\ \hline
 Ru & HfRu(B2) & HfRu(B2) & -819  \\ 
    & HfRu$_2$ \question &   &  \\ \hline
 Co & Co$_7$Hf$_2$(Ni$_7$Zr$_2$) &   & (-219) \\
    & Co$_2$Hf(C15) & Co$_2$Hf(C14) & -374[+3] \\
    & CoHf(B2) & CoHf(B33) & -401[+12] \\
    & CoHf$_2$(NiTi$_2$) & CoHf$_2$(C37) &  -314[+23]\\ \hline
 Ir & Hf$_2$Ir(NiTi$_2$) & Hf$_2$Ir(C37) & -750[+31]  \\
    & Hf$_5$Ir$_3$(D8$_8$ \cite{Massalski}, Ir$_5$Zr$_3$ \cite{Pauling}) & Hf$_5$Ir$_3$(Ir$_5$Zr$_3$) & -814[+14] \\
    & HfIr \question & HfIr(B27) & -949 \\
    &  & HfIr$_2$(Ga$_2$Hf) & -872 \\
    & HfIr$_3$(L1$_2$) & HfIr$_3$(L1$_2$) & -800 \\ \hline
 Rh & Hf$_2$Rh(NiTi$_2$) & Hf$_2$Rh(CuZr$_2$) & -632[+12] \\
    & HfRh(B2) & HfRh(B27) & -899[+30] \\
    & Hf$_3$Rh$_5$(Ge$_3$Rh$_5$) & Hf$_3$Rh$_5$(Ge$_3$Rh$_5$) & -928 \\
    & HfRh$_3$(L1$_2$) & HfRh$_3$(L1$_2$) & -762 \\   \hline
 Ni & Hf$_2$Ni(C16)& & (-345) \\
    & HfNi(CrB \cite{Massalski}, TlI \cite{Pauling}) & HfNi(TlI)& -542[+20] \\
    & Hf$_9$Ni$_{11}$(Zr$_9$Pt$_{11}$) &   & (-461) \\
    & Hf$_7$Ni$_{10}$(Zr$_7$Ni$_{10}$) &   & (-520) \\
    & HfNi$_3$(BaPb$_3$) & HfNi$_3$(D0$_{24}$) & -545[+4] \\
    & Hf$_2$Ni$_7$(Ni$_7$Zr$_2$)  &   & (-470) \\
    & HfNi$_5$(C15$_b$)  &   & (-350) \\ \hline
 Pt & Hf$_2$Pt(NiTi$_2$) & Hf$_2$Pt(NiTi$_2$) & -786 \\
    & HfPt(B2, B33 \cite{Massalski}, TlI \cite{Pauling}) & HfPt(B33, TlI) & -1153[+165] \\
\end{tabular}
\end{table} 
\vspace{-1mm}

\begin{table}[thw]
\scriptsize
\begin{tabular}{c|cc|c}
%
    & HfPt$_3$(L1$_2$, D0$_{24}$) & HfPt$_3$(D0$_{24}$) & -1100[+3] \\
    &    & HfPt$_8$(Pt$_8$Ti) & -528 \\ \hline 
 Pd & Hf$_2$Pd(C11$_b$ \cite{Massalski}, CuZr$_2$ \cite{Pauling}) & Hf$_2$Pd(C11$_b$, CuZr$_2$) & -527\\
    & HfPd \question & HfPd(B33) & -682 \\
    & Hf$_3$Pd$_4$ \question &  &  \\
    & HfPd$_2$(C11$_b$) & HfPd$_2$(C11$_b$) & -817 \\
    & HfPd$_3$(D0$_{24}$, L1$_2$) & HfPd$_3$(D0$_{24}$) & -879[+11] \\
    &    & HfPd$_5$(HfPd$_5^{\star}$) & -635 \\
    &    & HfPd$_8$(Pt$_8$Ti) & -430 \\ \hline
 Au & Au$_5$Hf \cite{Massalski},Au$_{4.2}$Hf$_{0.8}$ \cite{Pauling}(D1$_a$) &  & (-407)\\
    & Au$_4$Hf(Au$_4$Zr,D1$_a$) & Au$_4$Hf(Au$_4$Zr) & -414\\
    & Au$_3$Hf(D0$_a$) & Au$_3$Hf(D0$_a$) & -483 \\
    & Au$_2$Hf(C11$_b$) & Au$_2$Hf(C11$_b$) & -565 \\
    &    & Au$_4$Hf$_3$(Cu$_4$Ti$_3$)& -563\\
    & Au$_{10}$Hf$_7$ \question &    & \\
    & AuHf(B11) & AuHf(B11) & -545\\
    & AuHf$_2$(C11$_b$ \cite{Massalski}, CuZr$_2$ \cite{Pauling}) & AuHf$_2$(C11$_b$, CuZr$_2$ ) & -440 \\ \hline
 Ag & AgHf(B11) & AgHf(B11) & -119 \\
    & AgHf$_2$(C11$_b$ \cite{Massalski}, CuZr$_2$ \cite{Pauling}) & AgHf$_2$(C11$_b$, CuZr$_2$ ) & -122 \\ \hline
 Cu &    & Cu$_5$Hf(C15$_b$) &  -129\\
    & Cu$_{51}$Hf$_{14}$(Ag$_{51}$Gd$_{14}$) &  &  (139) \\   
    & Cu$_8$Hf$_3$(Cu$_8$Hf$_3$) & Cu$_8$Hf$_3$(Cu$_8$Hf$_3$) &  -173 \\
    & Cu$_{10}$Hf$_7$(Ni$_{10}$Zr$_7$) & Cu$_{10}$Hf$_7$(Ni$_{10}$Zr$_7$) & -186 \\
    & CuHf$_2$(C11$_b$ \cite{Massalski}, CuZr$_2$ \cite{Pauling}) & CuHf$_2$(C11$_b$, CuZr$_2$ ) & -166 \\ \hline
 Mg &  -  & HfMg(CdTi) & -7 \\ \hline
 Hg & Hf$_2$Hg(C11$_b$ \cite{Massalski}, CuZr$_2$ \cite{Pauling}) & Hf$_2$Hg(C11$_b$, CuZr$_2$ ) & -120 \\ \hline
 Cd & CdHf(B11 \cite{Massalski}, CdTi \cite{Pauling}) & CdHf(CdTi, B11)  & -88 \\
    & CdHf$_2$(C11$_b$ \cite{Massalski}, CuZr$_2$ \cite{Pauling}) & CdHf$_2$($\beta1$) & -87[+18] \\ \hline
 Zn & Hf$_2$Zn(C11$_b$ \cite{Massalski}, CuZr$_2$ \cite{Pauling}) & Hf$_2$Zn(C11$_b$, CuZr$_2$) & -175 \\
    & HfZn$_2$(C36 \cite{Massalski}, C15 \cite{Pauling}) & HfZn$_2$(CaIn$_2$) & -233[+18] \\
    & HfZn$_3$ \question & HfZn$_3$(YCd$_3$) & -204 \\
    & HfZn$_5$ \question &  & \\
    & HfZn$_{22}$(Zn$_{22}$Zr) & HfZn$_{22}$(Zn$_{22}$Zr) & -43 \\   \hline 
 Be & Be$_{13}$Hf(D2$_3$)  & Be$_{13}$Hf(D2$_3$) & -173 \\ 
    & Be$_{17}$Hf$_2$(Th$_2$Ni$_{17}$,Th$_2$Zn$_{17}$) & Be$_{17}$Hf$_2$(Th$_2$Zn$_{17}$) & -223[+485] \\ 
    & Be$_5$Hf(D2$_d$) & Be$_5$Hf(D2$_d$) & -226 \\
    & Be$_2$Hf(C32)&  & (-166)\\ 
    & BeHf(TlI)&  & (-75)\\  \hline
 Tl &  -  & Hf$_2$Tl($\beta_2$) & -2 \\ \hline
 In &   & Hf$_3$In(L1$_2$)& -145 \\
    & HfIn(L1$_0$) &  & (-160) \\ 
    & Hf$_3$In$_4$(In$_4$Ti$_3$) & Hf$_3$In$_4$(In$_4$Ti$_3$) & -285 \\  \hline
 Al & Al$_3$Hf(D0$_{22}$ \cite{Massalski}, D0$_{23}$ \cite{Pauling}) & Al$_3$Hf(D0$_{23}$) & -356[+9]\\
    & Al$_2$Hf(C14) & Al$_2$Hf(C14) & -415\\
    & Al$_3$Hf$_2$(Al$_3$Zr$_2$) &   & (-398) \\
    & AlHf(B33 \cite{Massalski}, TlI \cite{Pauling}) &   & (-380) \\
    & Al$_3$Hf$_4$(Al$_3$Zr$_4$) & Al$_3$Hf$_4$(Al$_3$Zr$_4$) & -374 \\ 
    & Al$_2$Hf$_3$(Al$_2$Zr$_3$) &   & (-314) \\
    & Al$_3$Hf$_5$(D8$_8$) &   & (-277) \\ 
    & AlHf$_2$(C16)  &  & (-252)\\ 
    & AlHf$_3$ \question \cite{Pauling} & AlHf$_3$(L1$_2$) & -224\\ \hline
 Ga & Ga$_3$Hf(D0$_{22}$) &   & (-382) \\
    & Ga$_2$Hf(Ga$_2$Hf)  &   & (-456) \\
    & Ga$_3$Hf$_2$(Al$_3$Zr$_2$) & Ga$_3$Hf$_2$(Al$_3$Zr$_2$) & -664 \\
    & GaHf(InTh)  &   & (-519) \\ 
    & Ga$_{10}$Hf$_{11}$(Ge$_{10}$Ho$_{11}$) &   & (-495) \\
    & Ga$_3$Hf$_5$(D8$_8$) & Ga$_3$Hf$_5$(D8$_8$) &  -445 \\
    & GaHf$_2$(C16) & GaHf$_2$(C16) & -404 \\ \hline
 Pb &  -  & Hf$_5$Pb$^\star$ & -140 \\
    &   & Hf$_5$Pb$_3$(W$_5$Si$_3$) &  -253 \\ \hline
 Sn & Hf$_5$Sn$_3$(D8$_8$) & Hf$_5$Sn$_3$(D8$_8$) & -380 \\
    & Hf$_5$Sn$_4$(Ga$_4$Ti$_5$) & Hf$_5$Sn$_4$(Ga$_4$Ti$_5$) & -396 \\
    & HfSn(B20) &   & (-270) \\
    & HfSn$_2$(C40) & HfSn$_2$(C40) & -265 \\ \hline
 Bi & Bi$_2$Hf(As$_2$Ti) & Bi$_2$Hf(C16) & -147[+14]\\
    & Bi$_9$Hf$_8$(V$_{7.46}$Sb$_9$) &  & (-166) \\
    & BiHf \question & BiHf(B11) & -182 \\ 
    &        & BiHf$_2^\star$ & -169\\ \hline
\hline
\end{tabular}
\end{table} 


\begin{widetext}{

\begin{figure}[htb]
  \includegraphics[width=\figswidth]{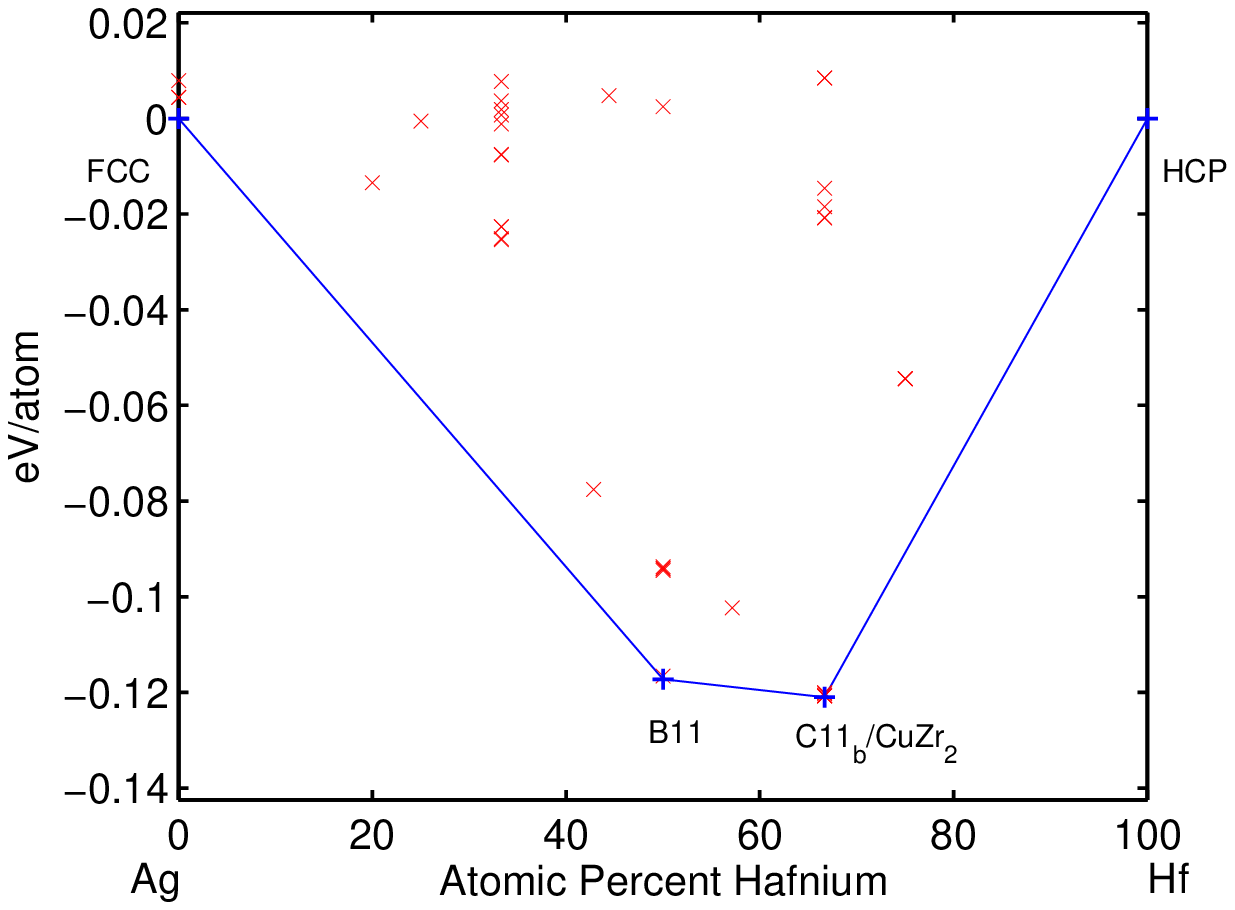}\hspace{\figsskip}
  \includegraphics[width=\figswidth]{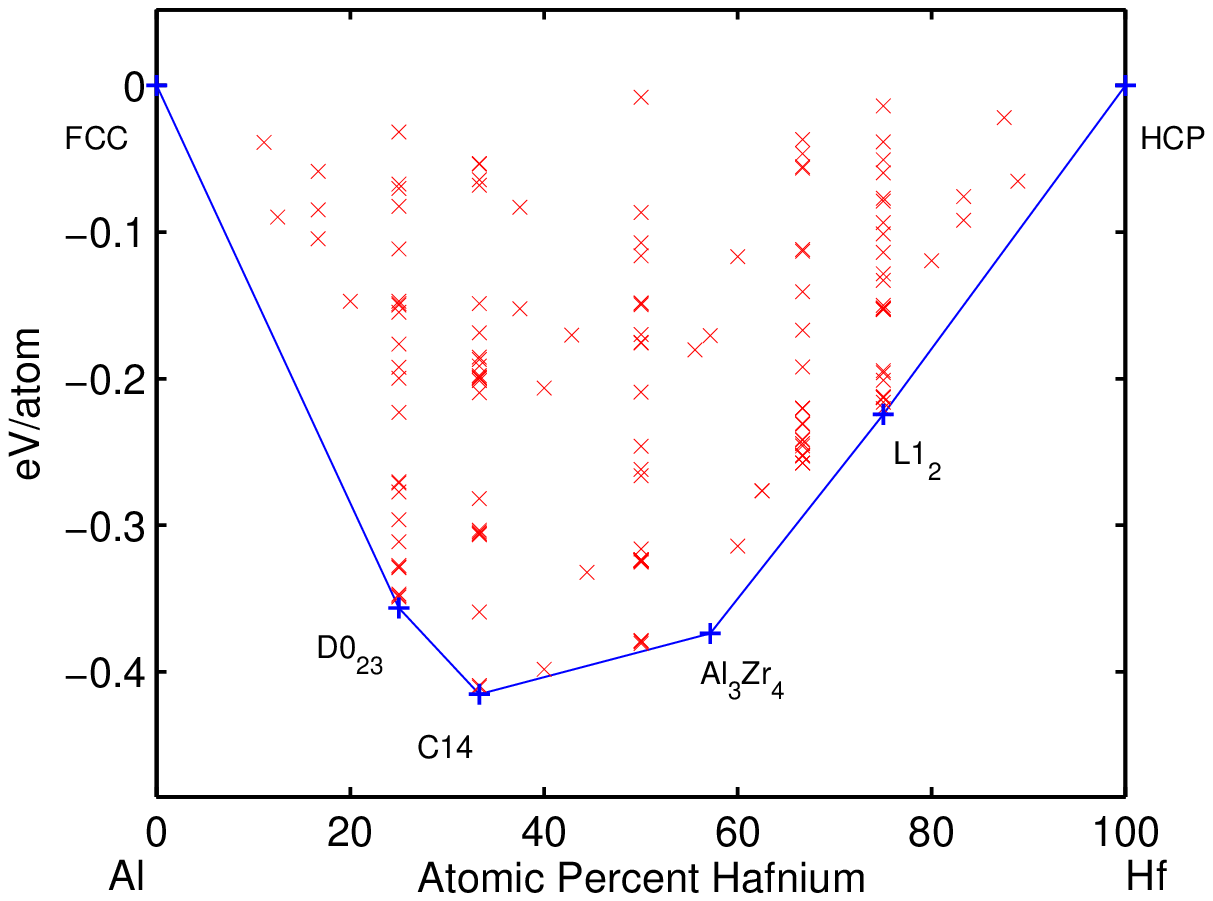}\hspace{\figsskip}
  \includegraphics[width=\figswidth]{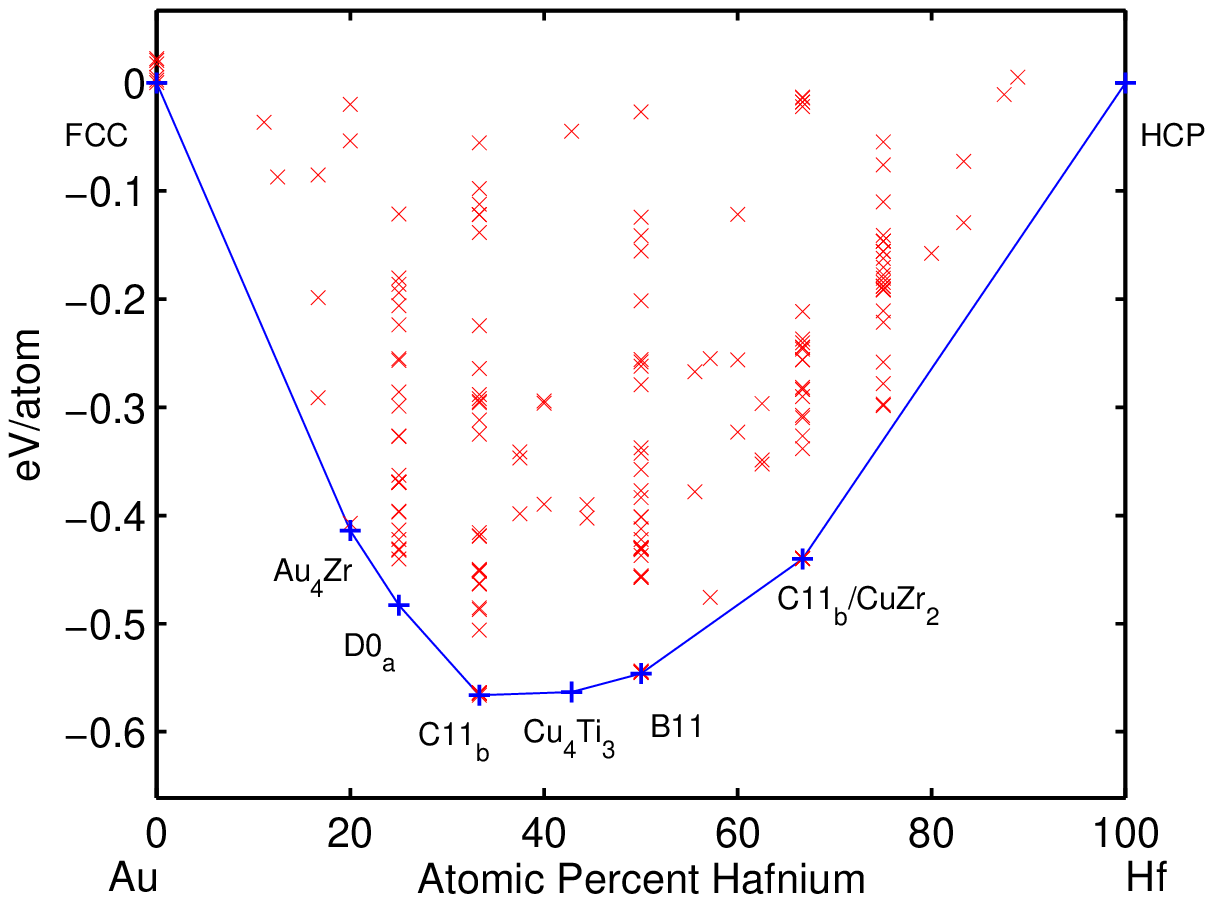}\hspace{\figsskip}
  \includegraphics[width=\figswidth]{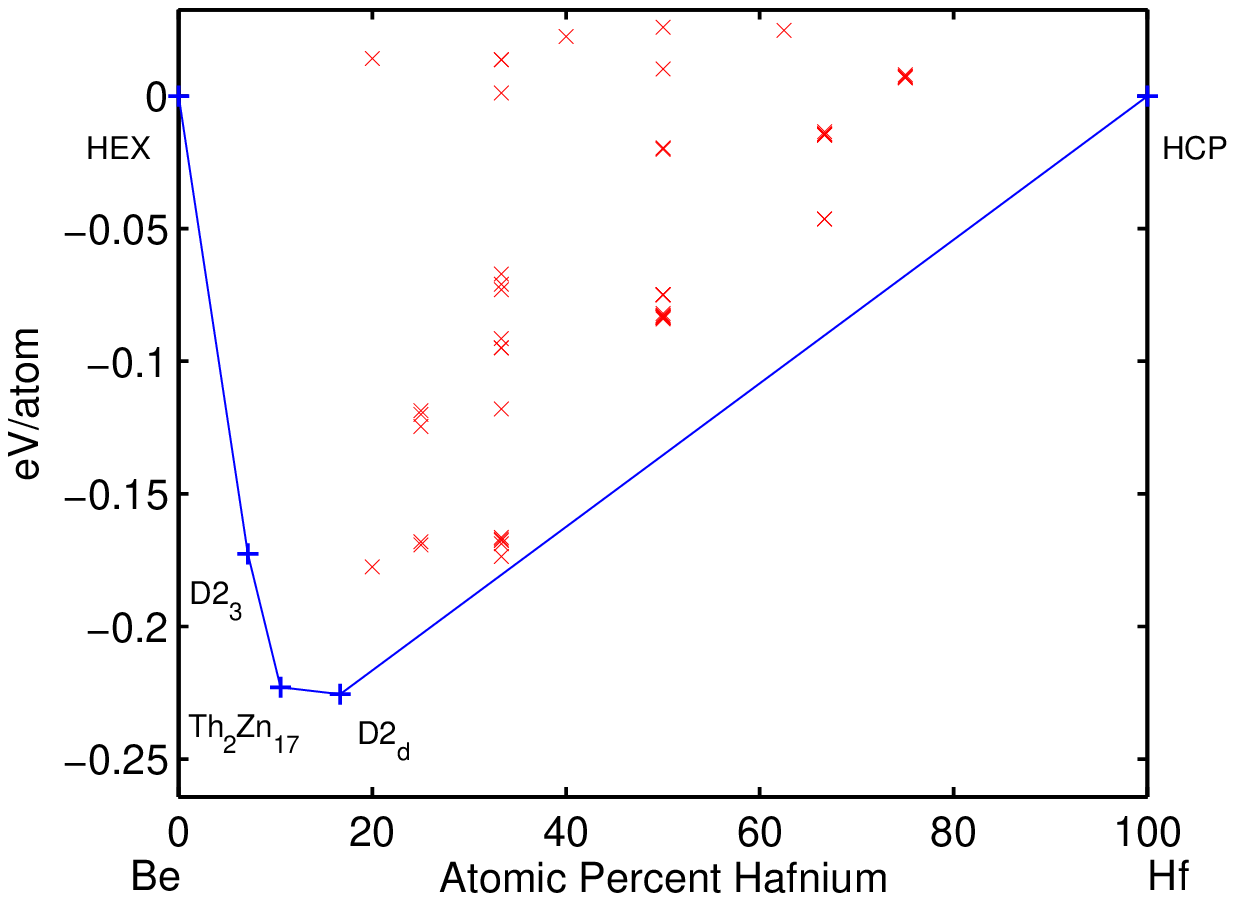}\hspace{\figsskip}
  \includegraphics[width=\figswidth]{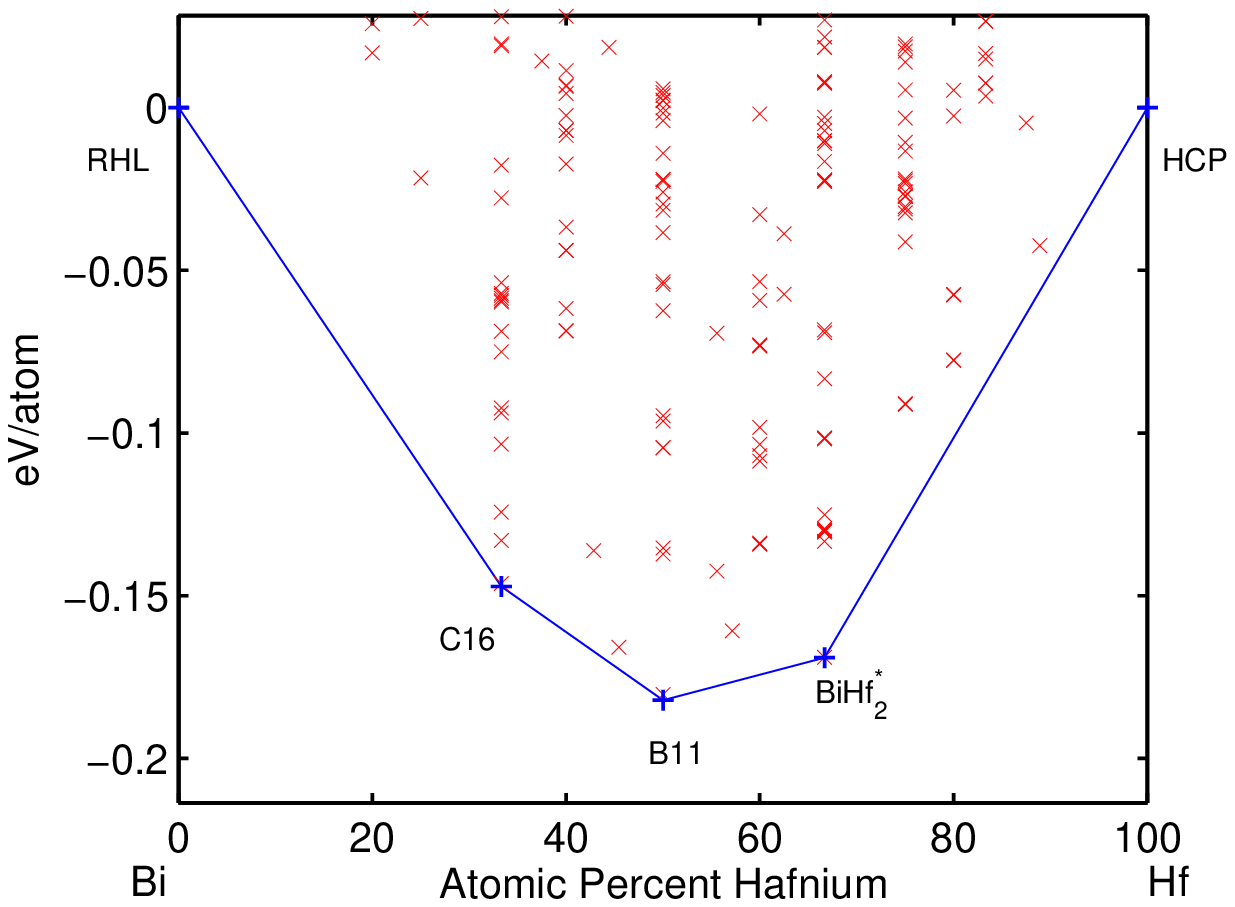}\hspace{\figsskip}
  \includegraphics[width=\figswidth]{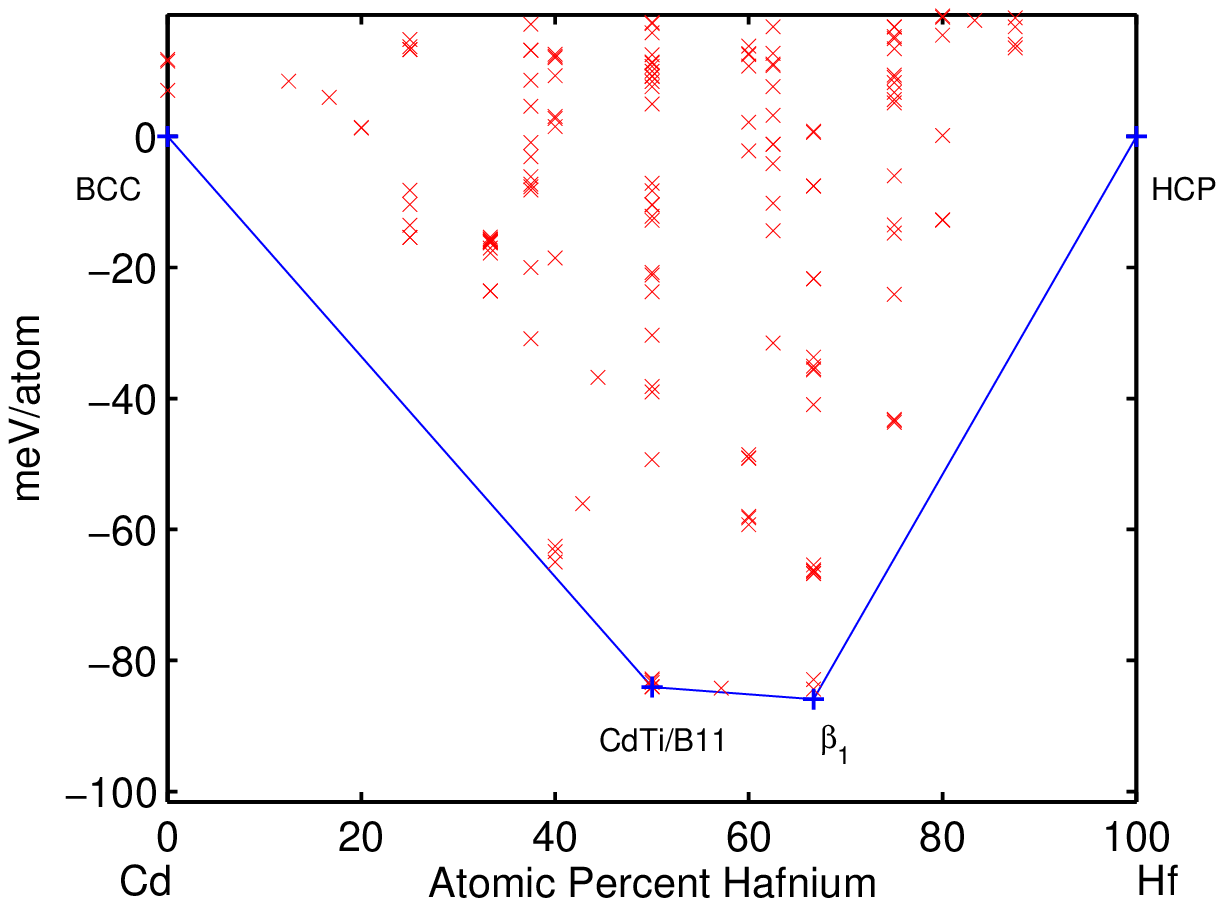}\hspace{\figsskip}
  \includegraphics[width=\figswidth]{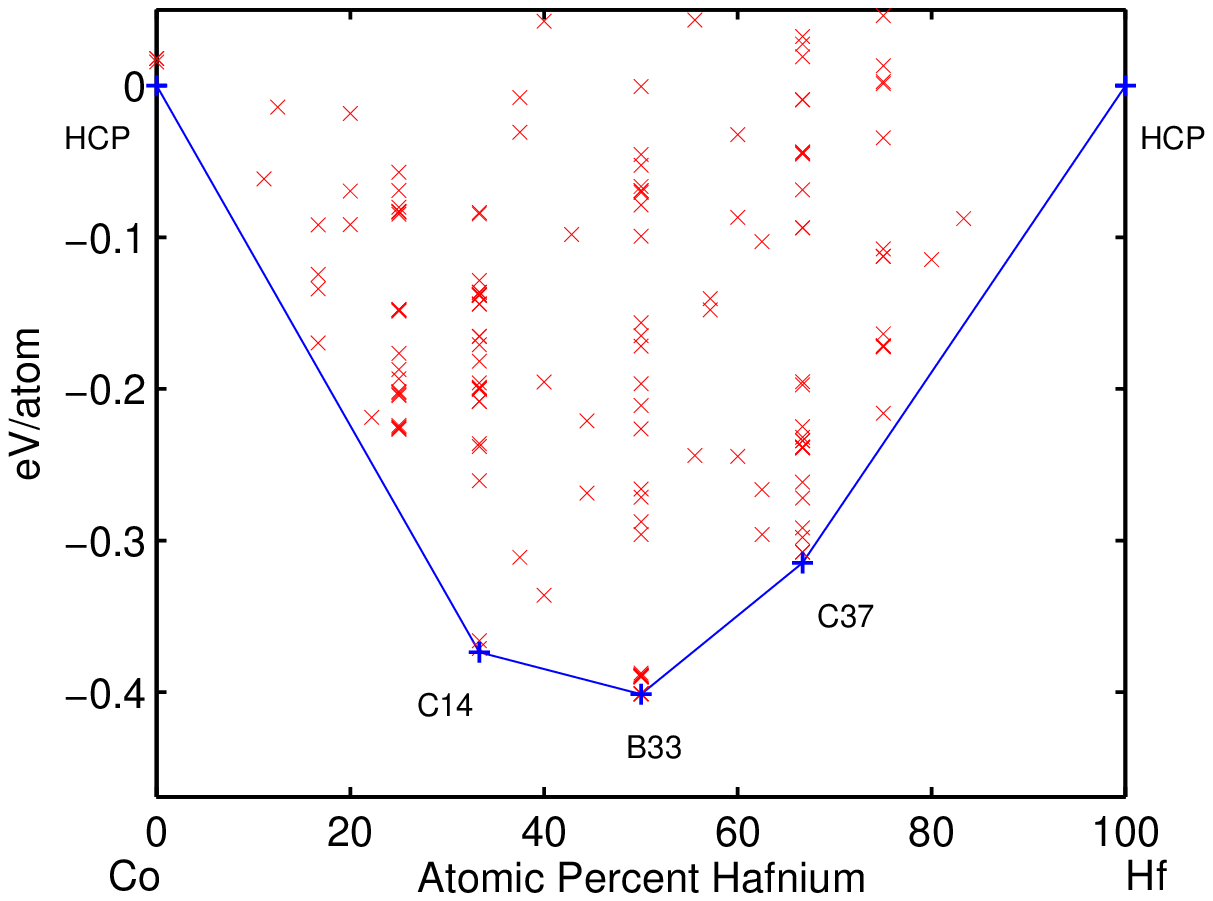}\hspace{\figsskip}
  \includegraphics[width=\figswidth]{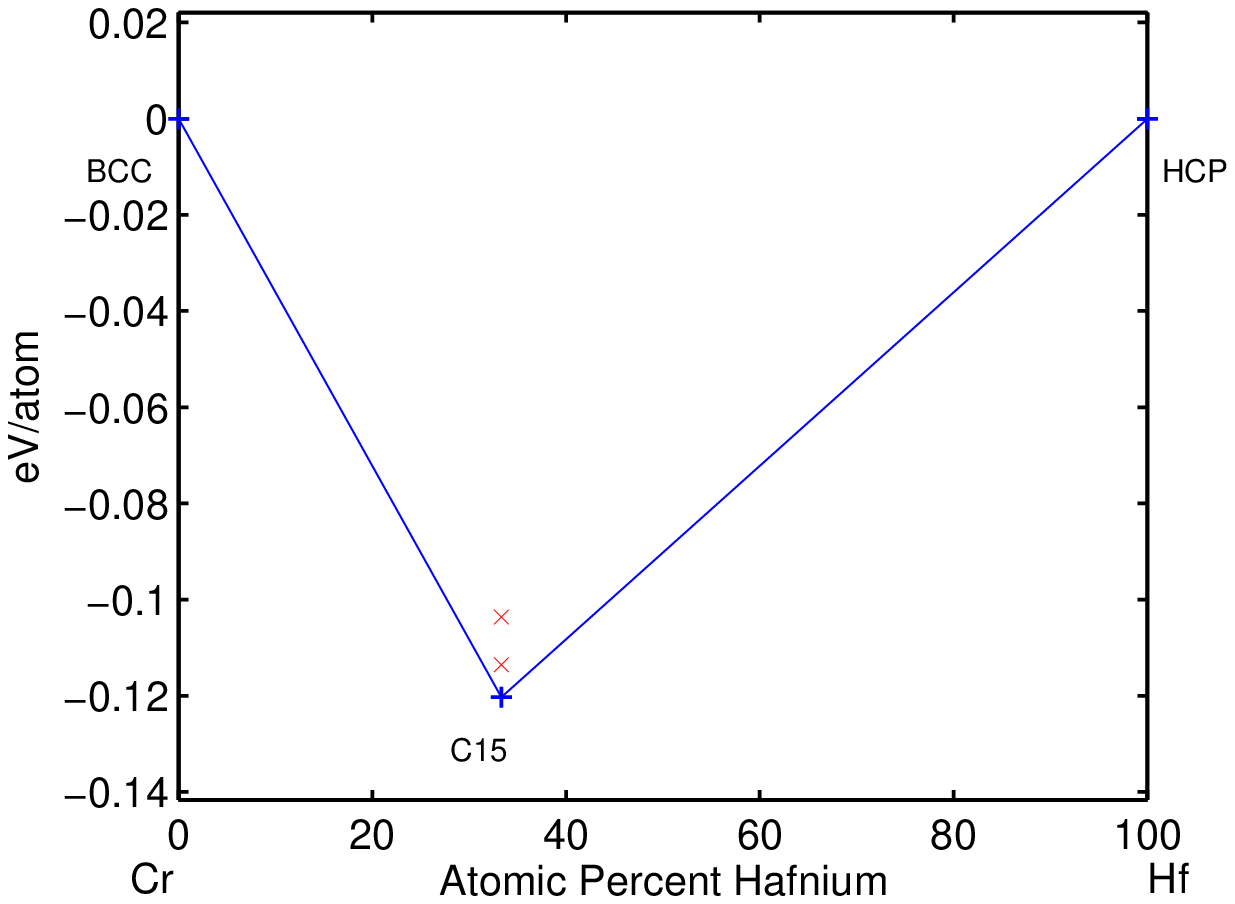}\hspace{\figsskip}
  \includegraphics[width=\figswidth]{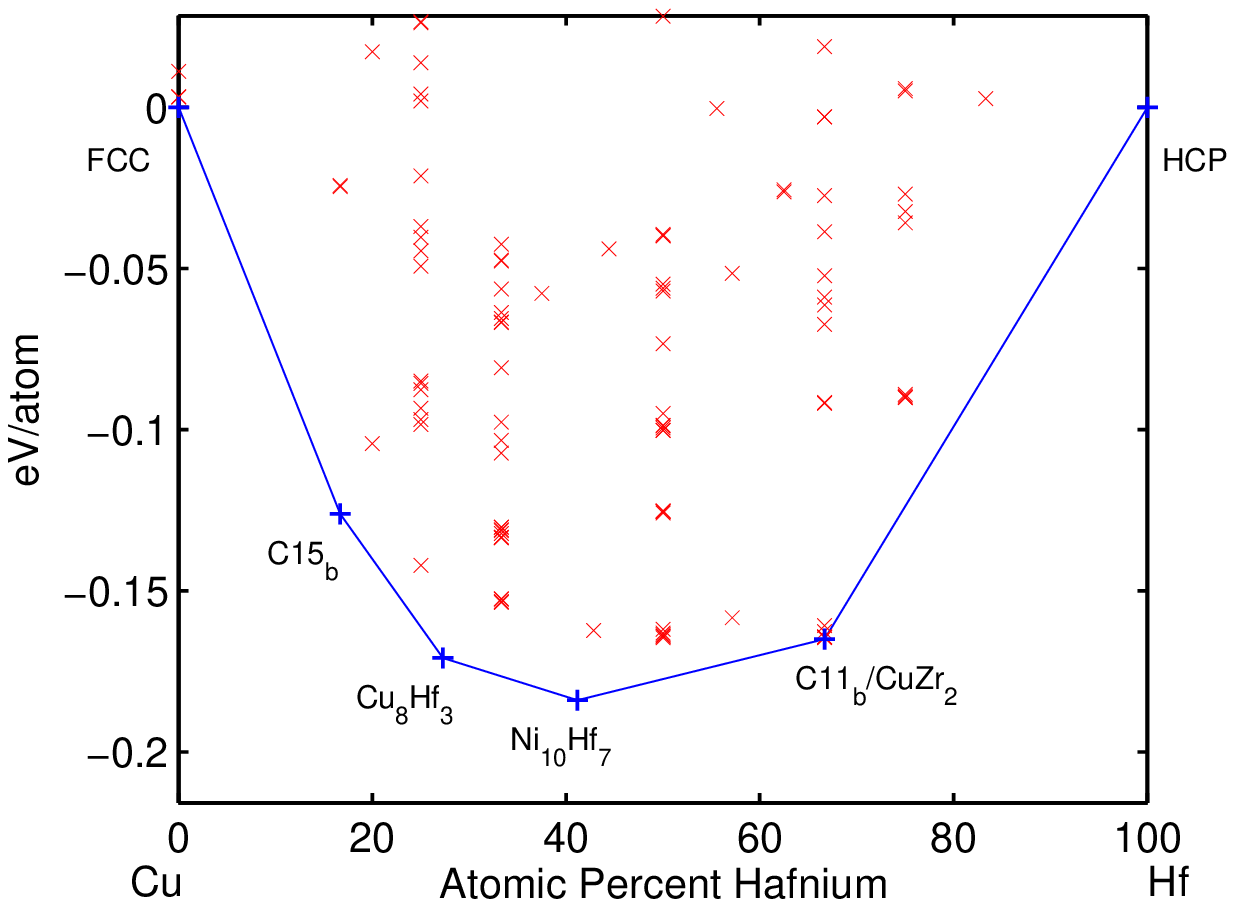}\hspace{\figsskip}
  \includegraphics[width=\figswidth]{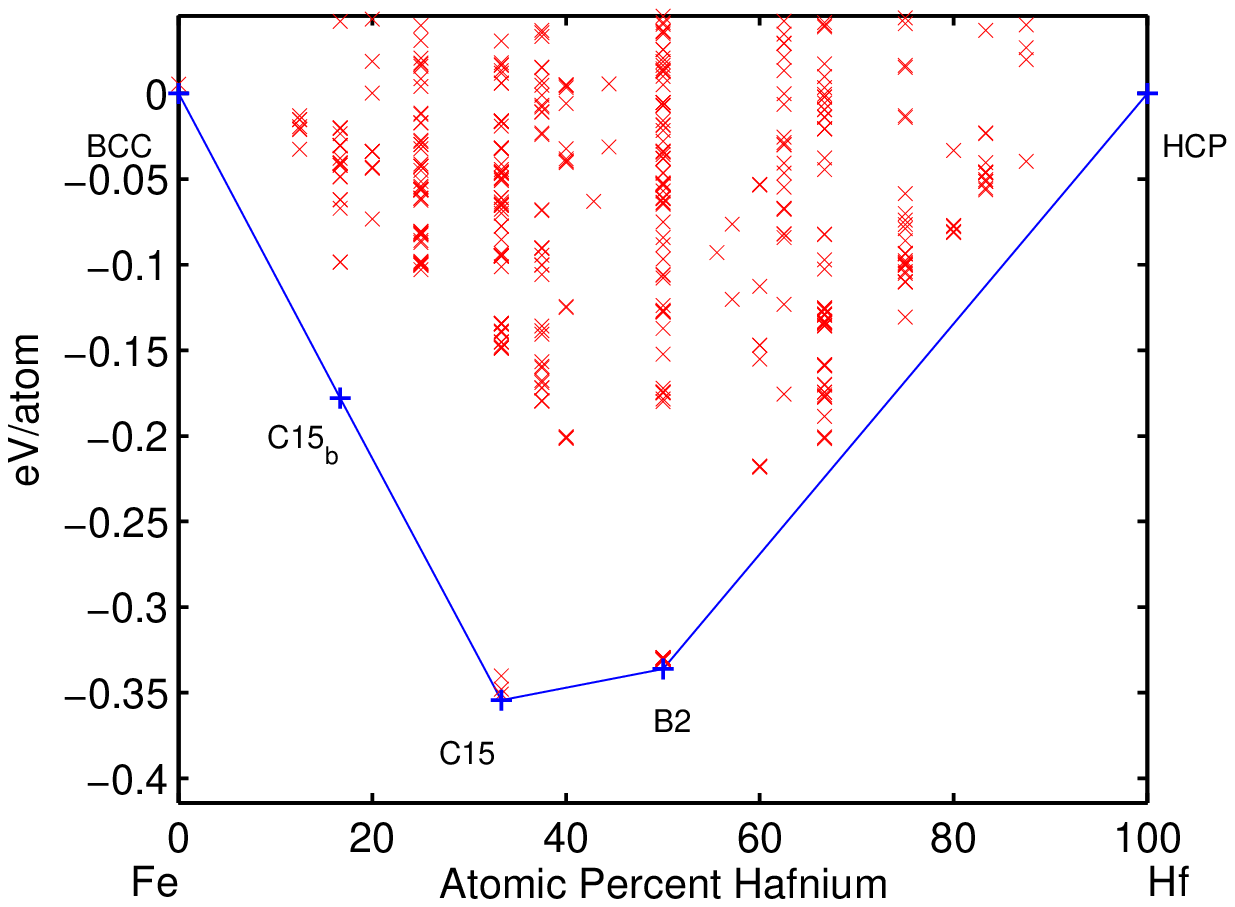}\hspace{\figsskip}
  \includegraphics[width=\figswidth]{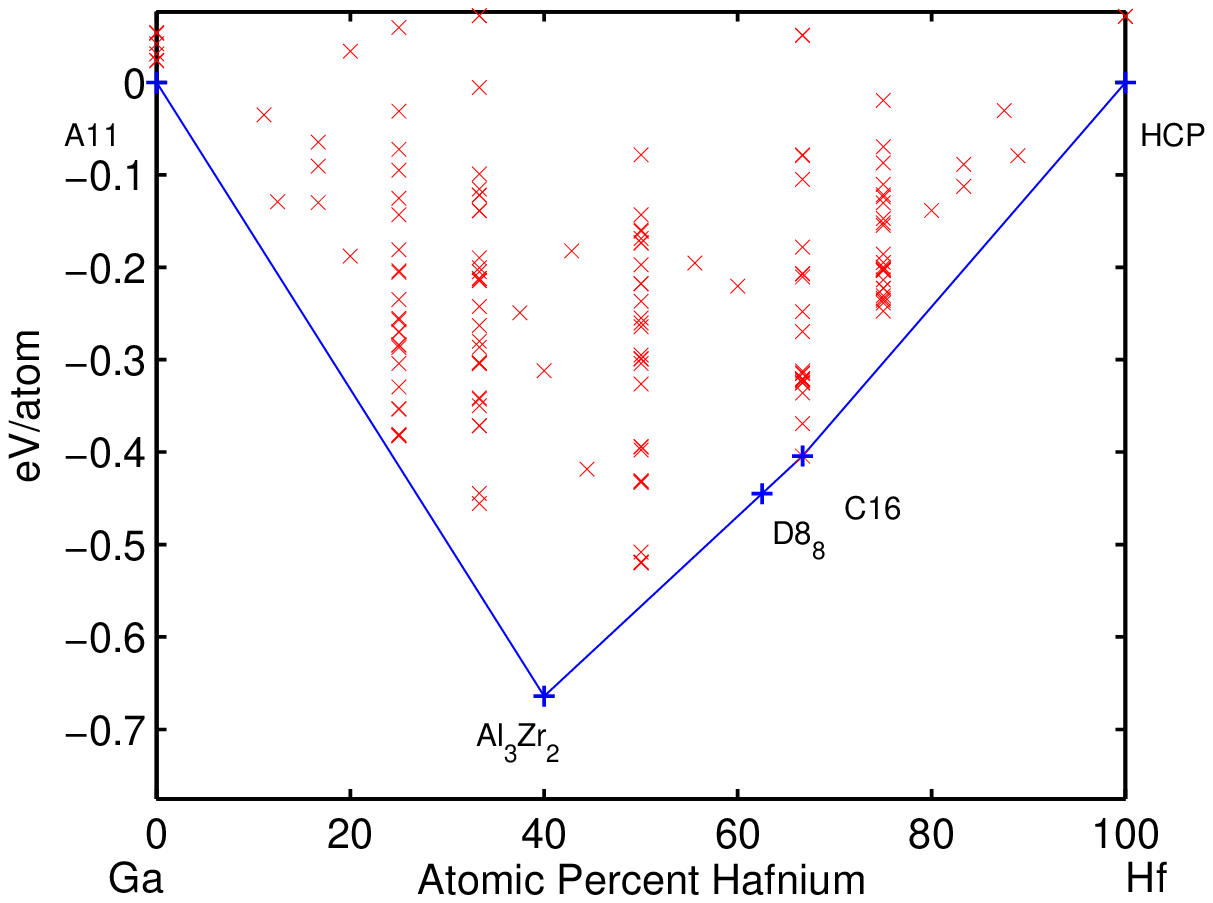}\hspace{\figsskip}
  \includegraphics[width=\figswidth]{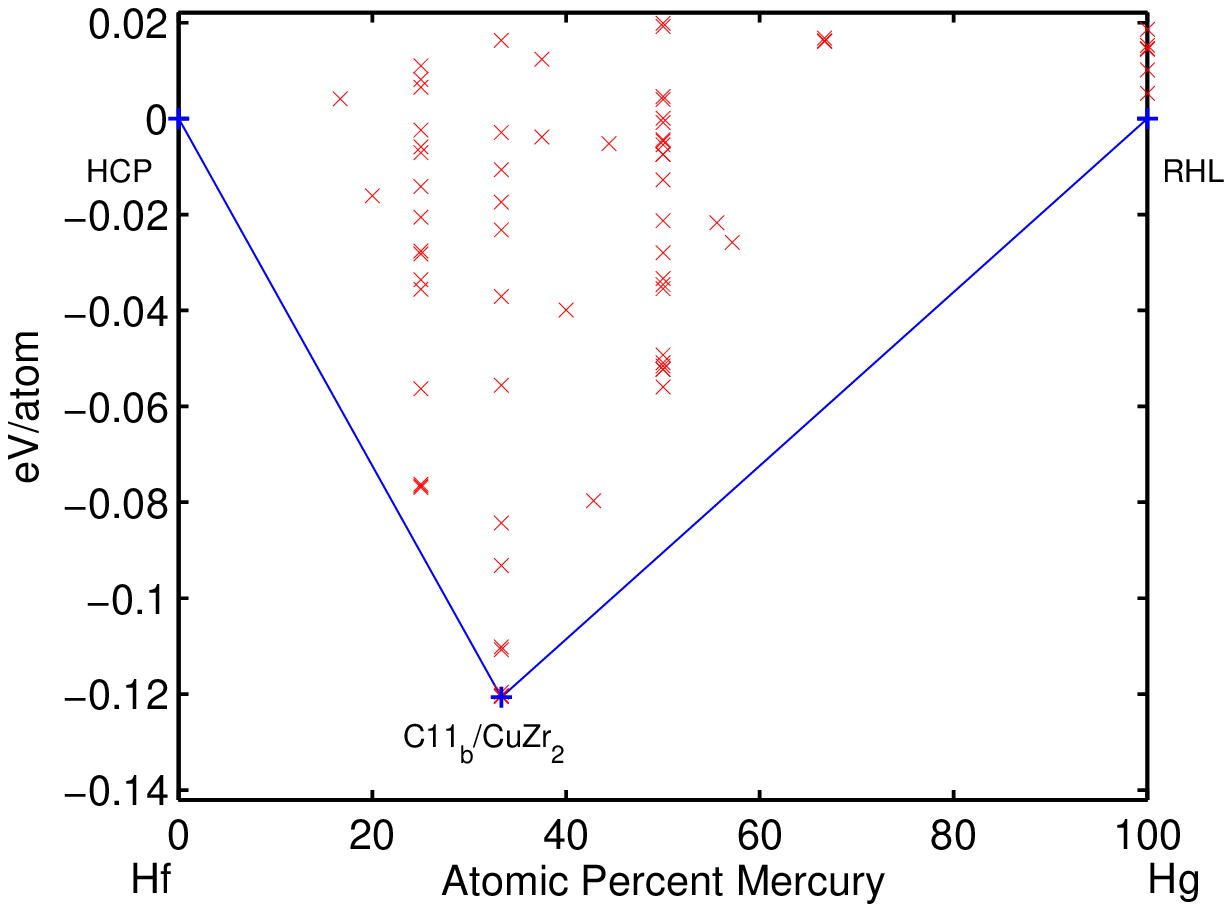}\hspace{\figsskip}
  \includegraphics[width=\figswidth]{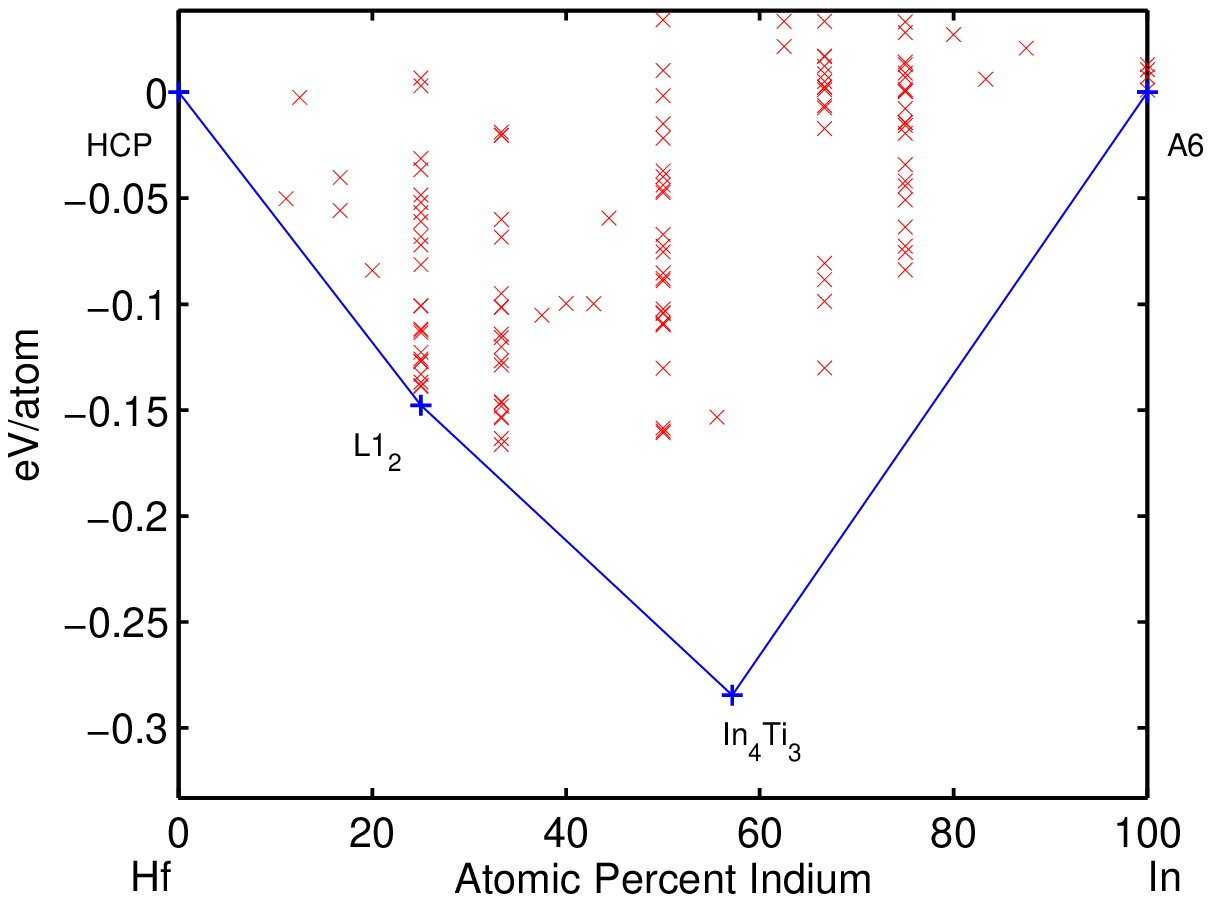}\hspace{\figsskip}
  \includegraphics[width=\figswidth]{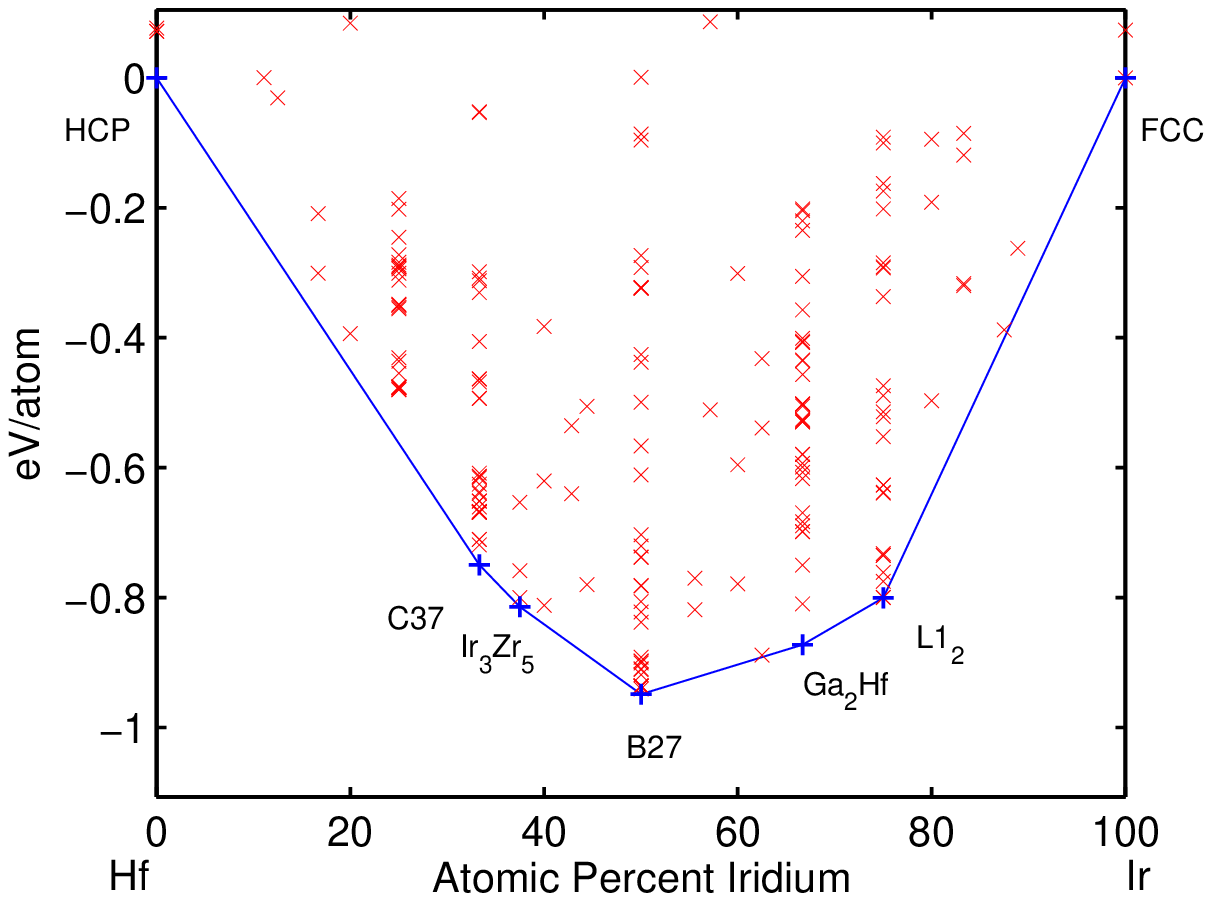}\hspace{\figsskip}
  \includegraphics[width=\figswidth]{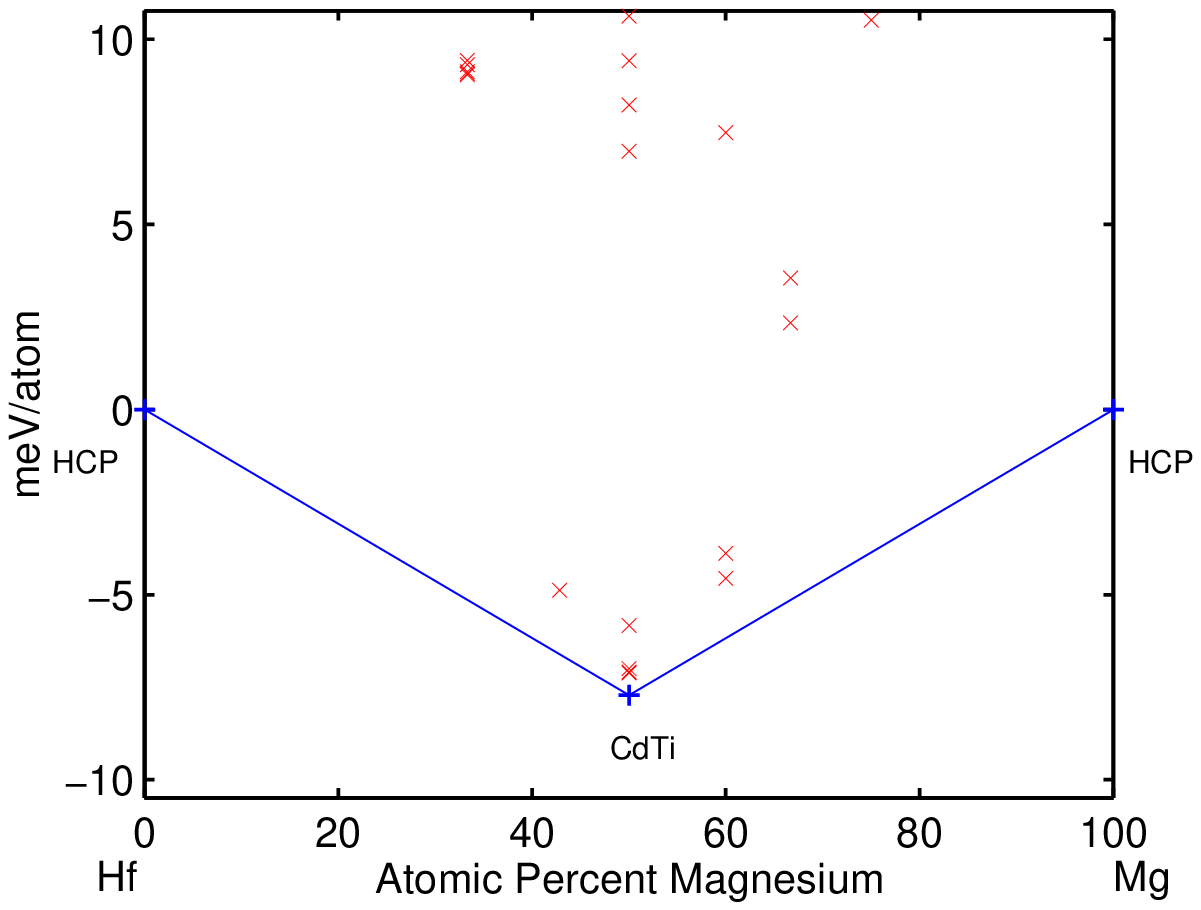}\hspace{\figsskip}
  \vnegspace
  \caption{{\small Formation enthalpies of
      Ag-Hf, Al-Hf, Au-Hf, 
      Be-Hf, Bi-Hf, Cd-Hf, 
      Co-Hf, Au-Hf, Cu-Hf, 
      Fe-Hf, Ga-Hf, Hf-Hg, 
      Hf-In, Hf-Ir, and Hf-Mg 
      alloys.}}
  \vnegspace
  \label{fig_AgHf}\label{fig_AlHf}\label{fig_AuHf}
  \label{fig_BeHf}\label{fig_BiHf}\label{fig_CdHf}
  \label{fig_CoHf}\label{fig_CrHf}\label{fig_CuHf}
  \label{fig_FeHf}\label{fig_GaHf}\label{fig_HfHg}
  \label{fig_HfIn}\label{fig_HfIr}\label{fig_HfMg}
\end{figure}
}\end{widetext}

\newpage

\begin{widetext}{

\begin{figure}[htb]
  \includegraphics[width=\figswidth]{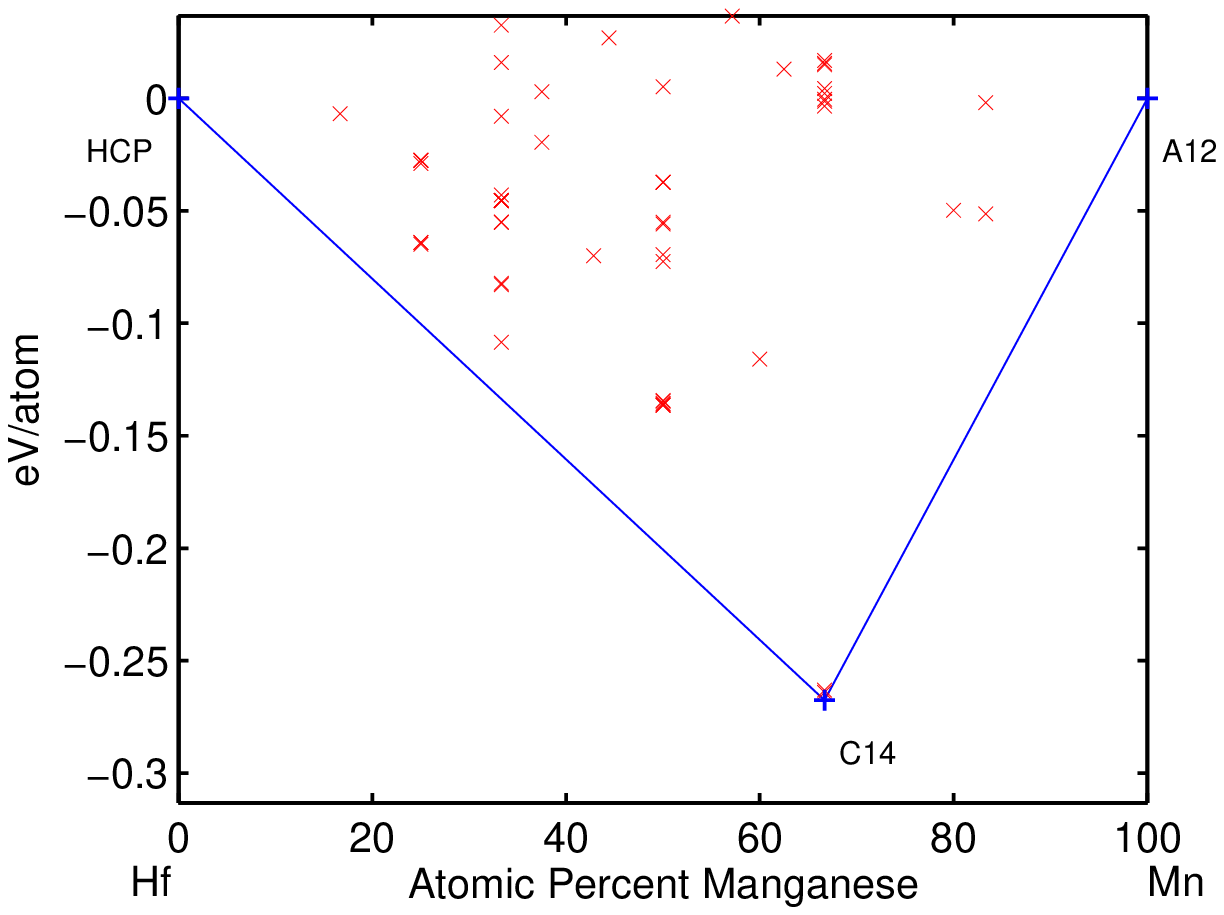}\hspace{\figsskip}
  \includegraphics[width=\figswidth]{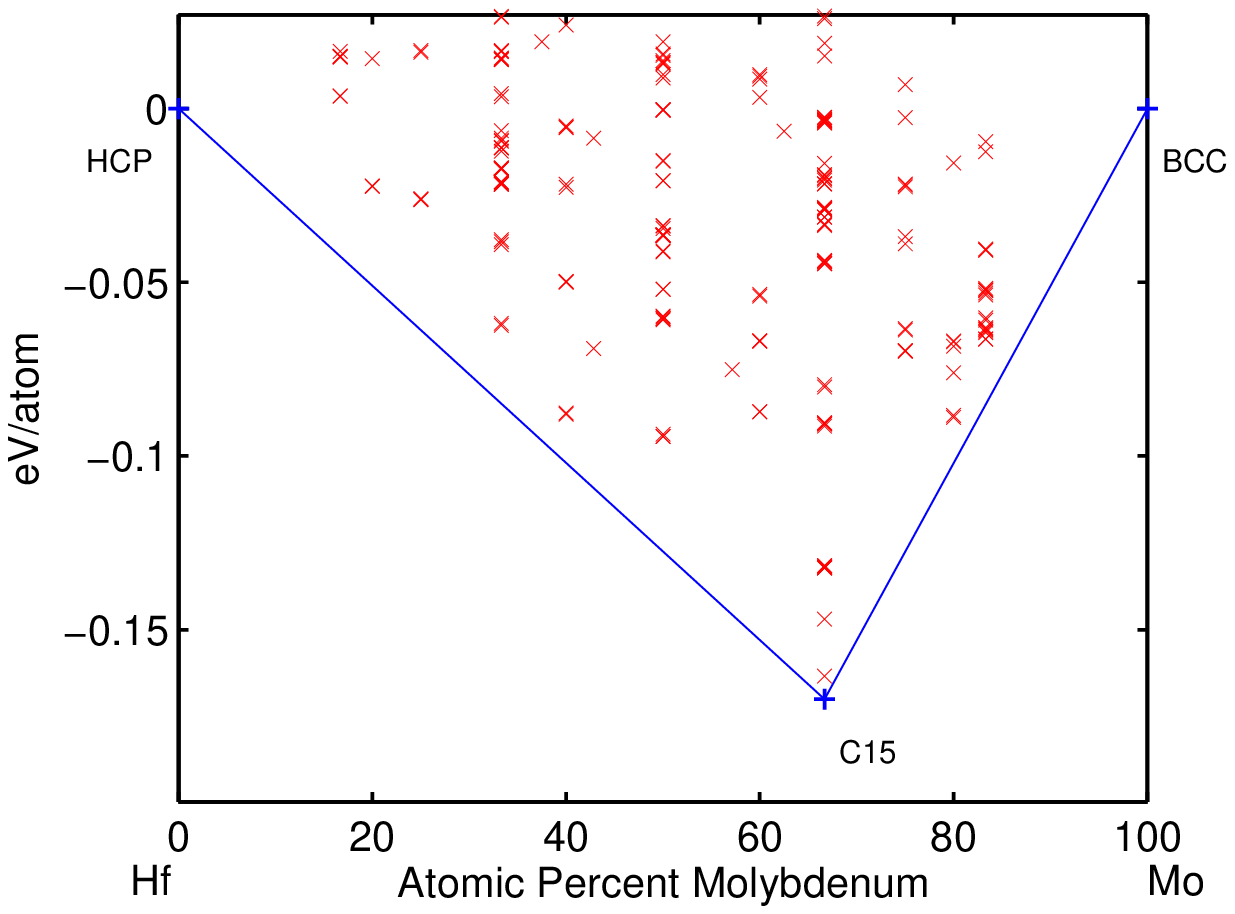}\hspace{\figsskip}
  \includegraphics[width=\figswidth]{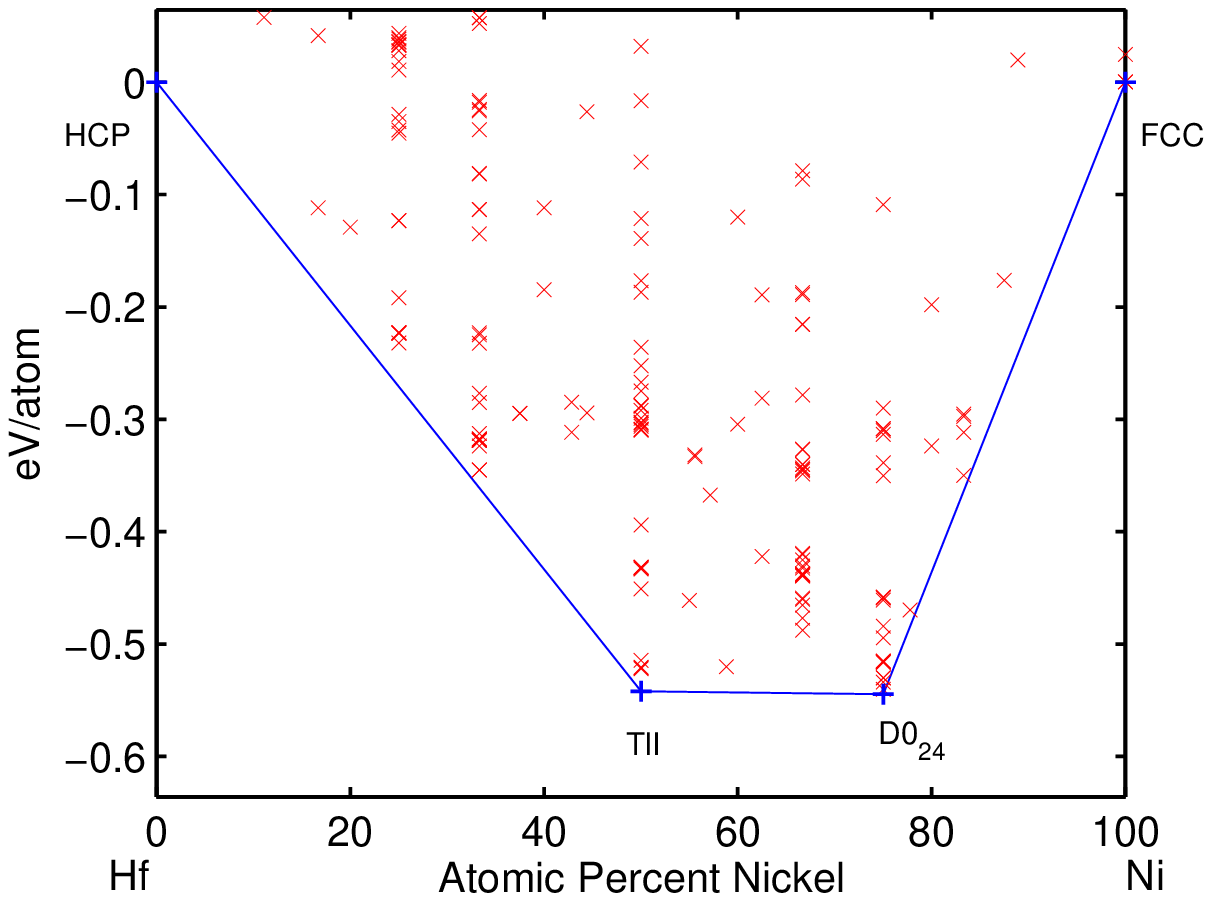}\hspace{\figsskip}
  \includegraphics[width=\figswidth]{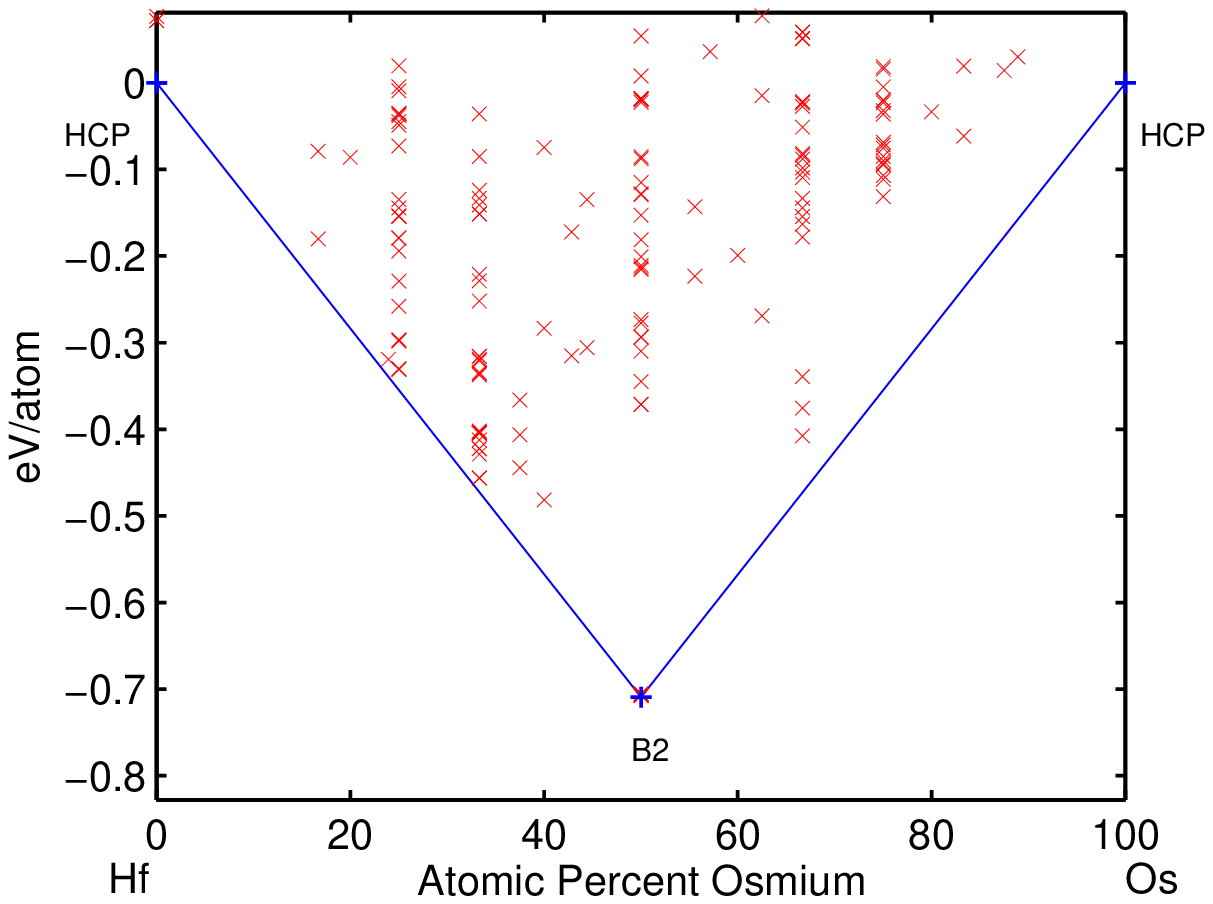}\hspace{\figsskip}
  \includegraphics[width=\figswidth]{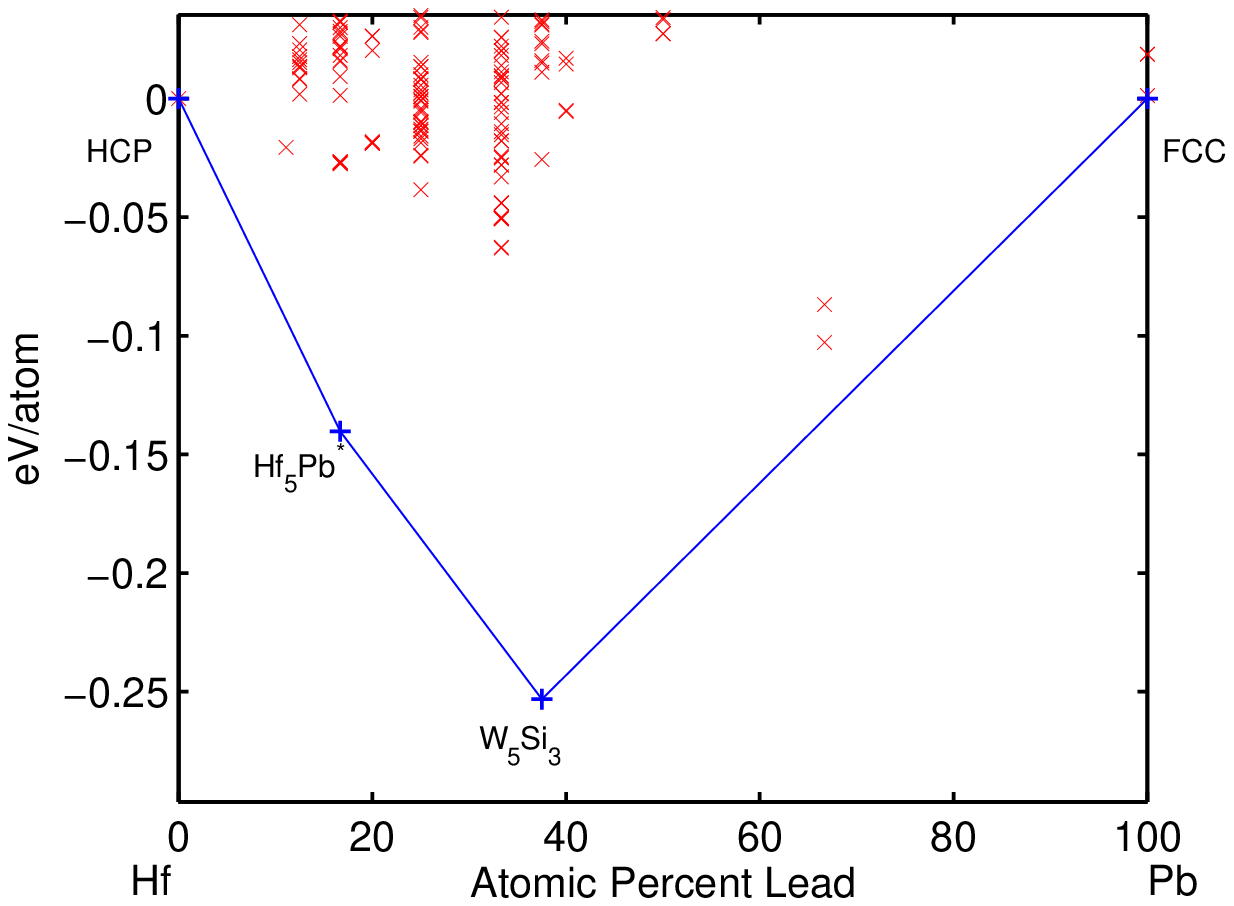}\hspace{\figsskip}
  \includegraphics[width=\figswidth]{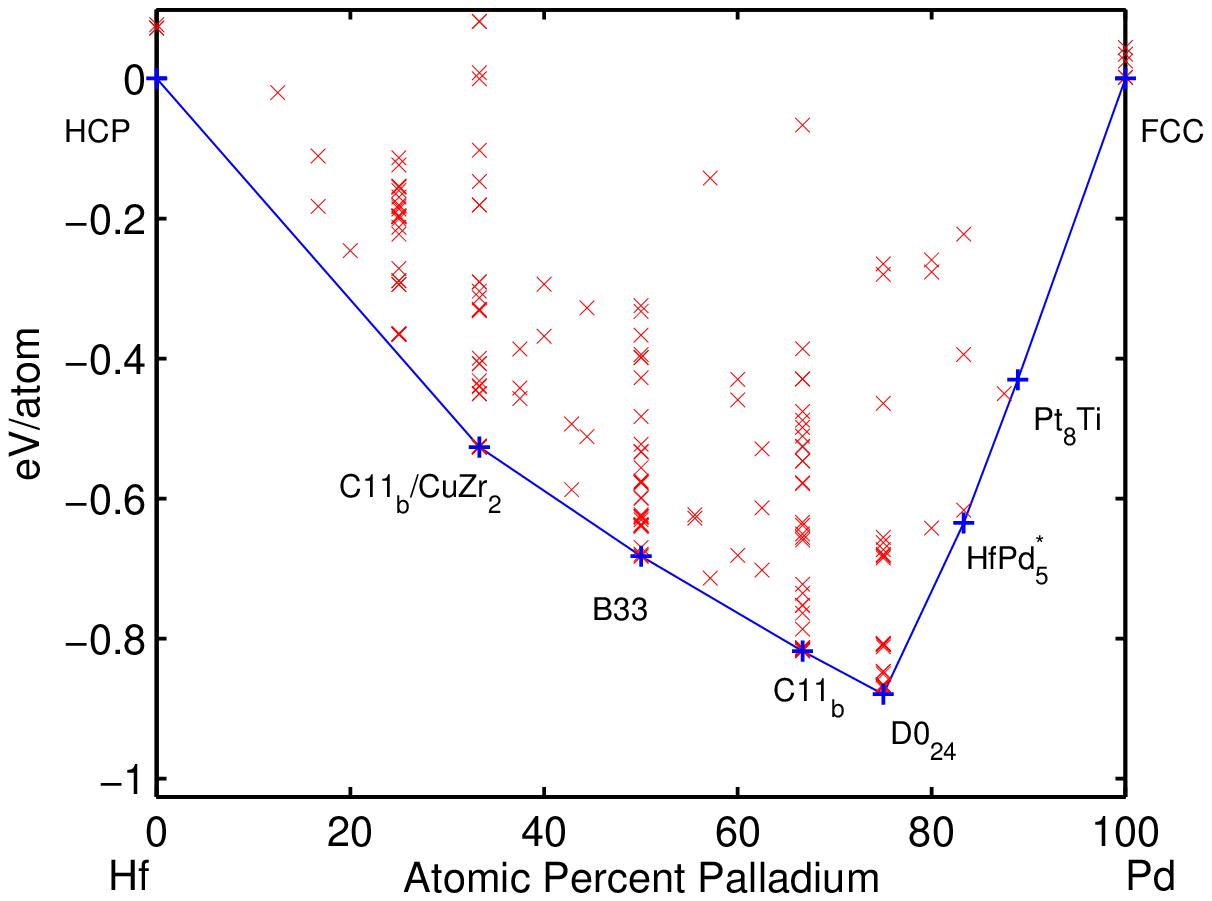}\hspace{\figsskip}
  \includegraphics[width=\figswidth]{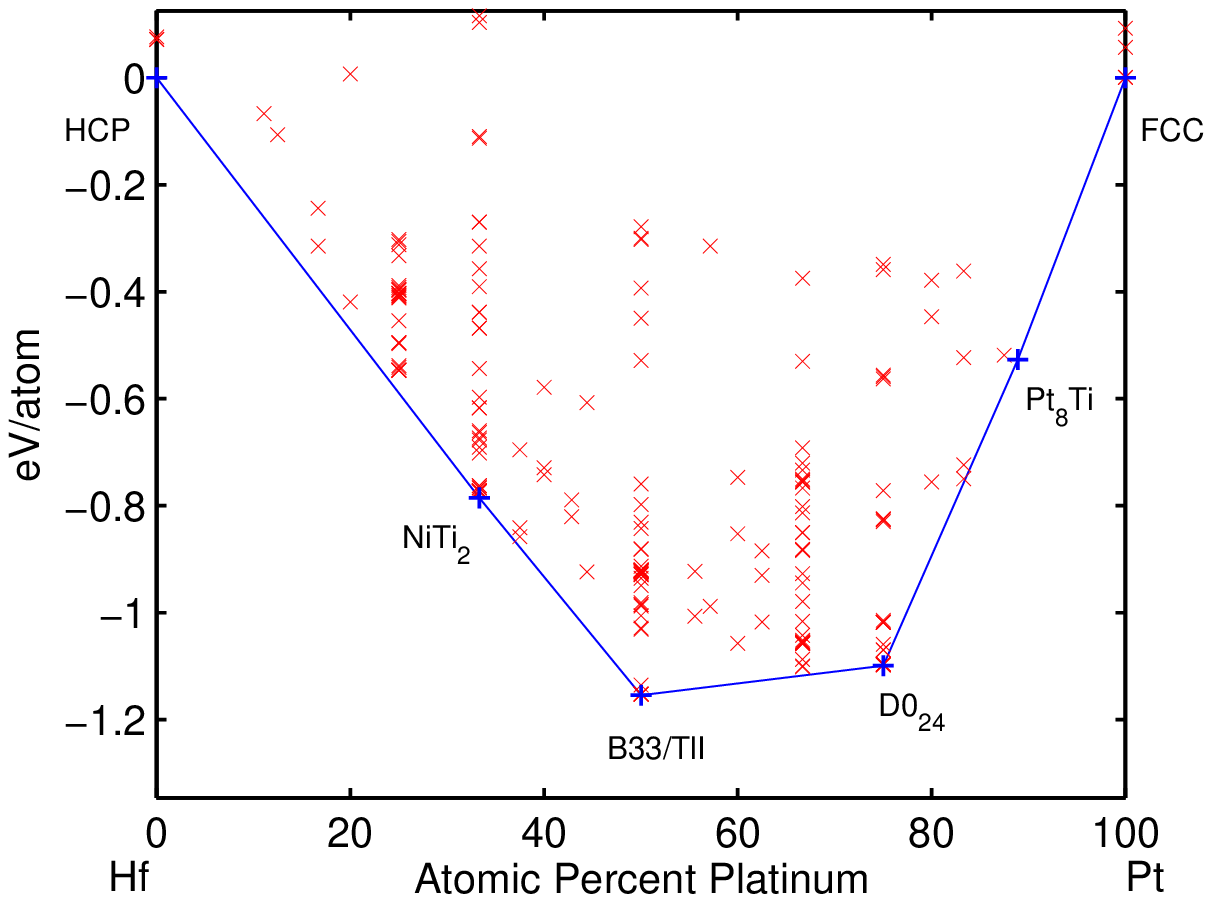}\hspace{\figsskip}
  \includegraphics[width=\figswidth]{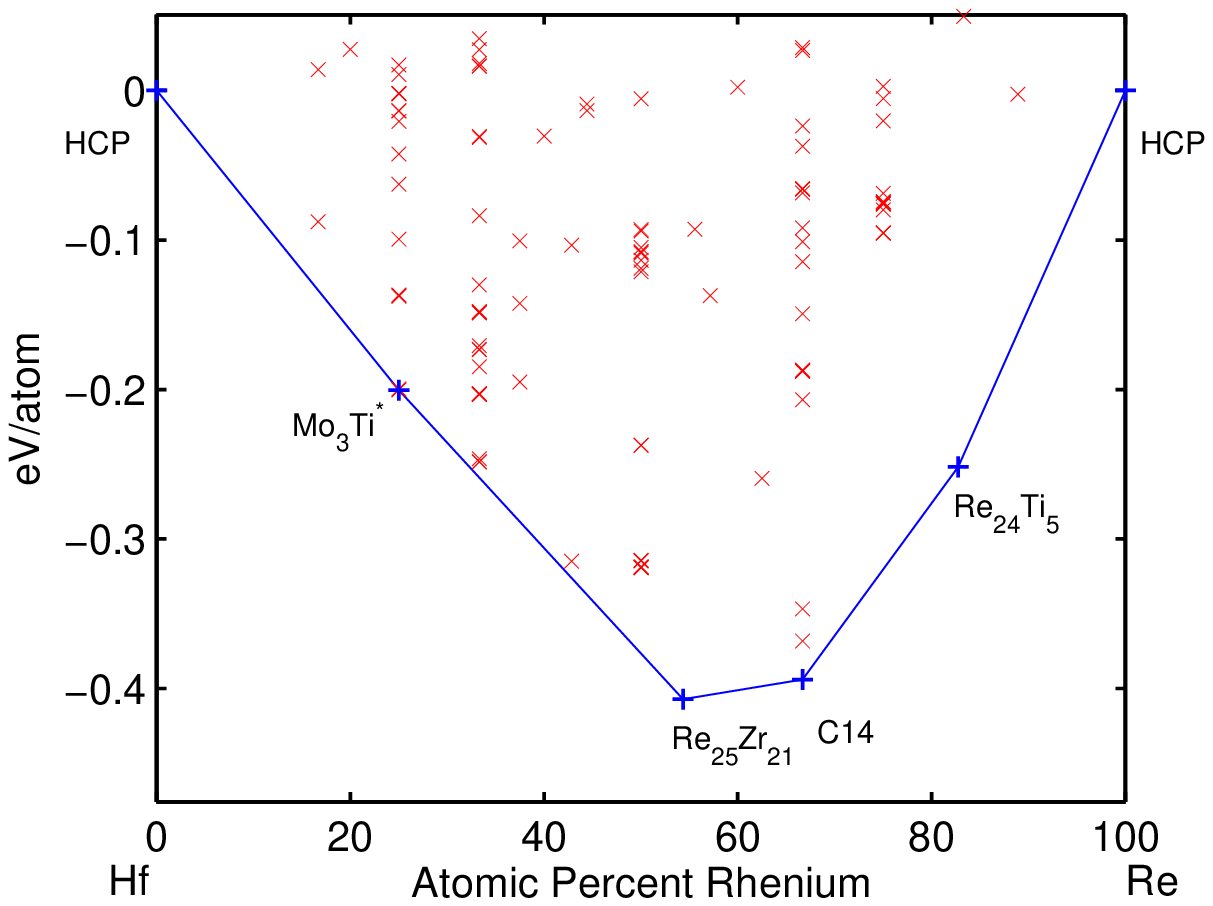}\hspace{\figsskip}
  \includegraphics[width=\figswidth]{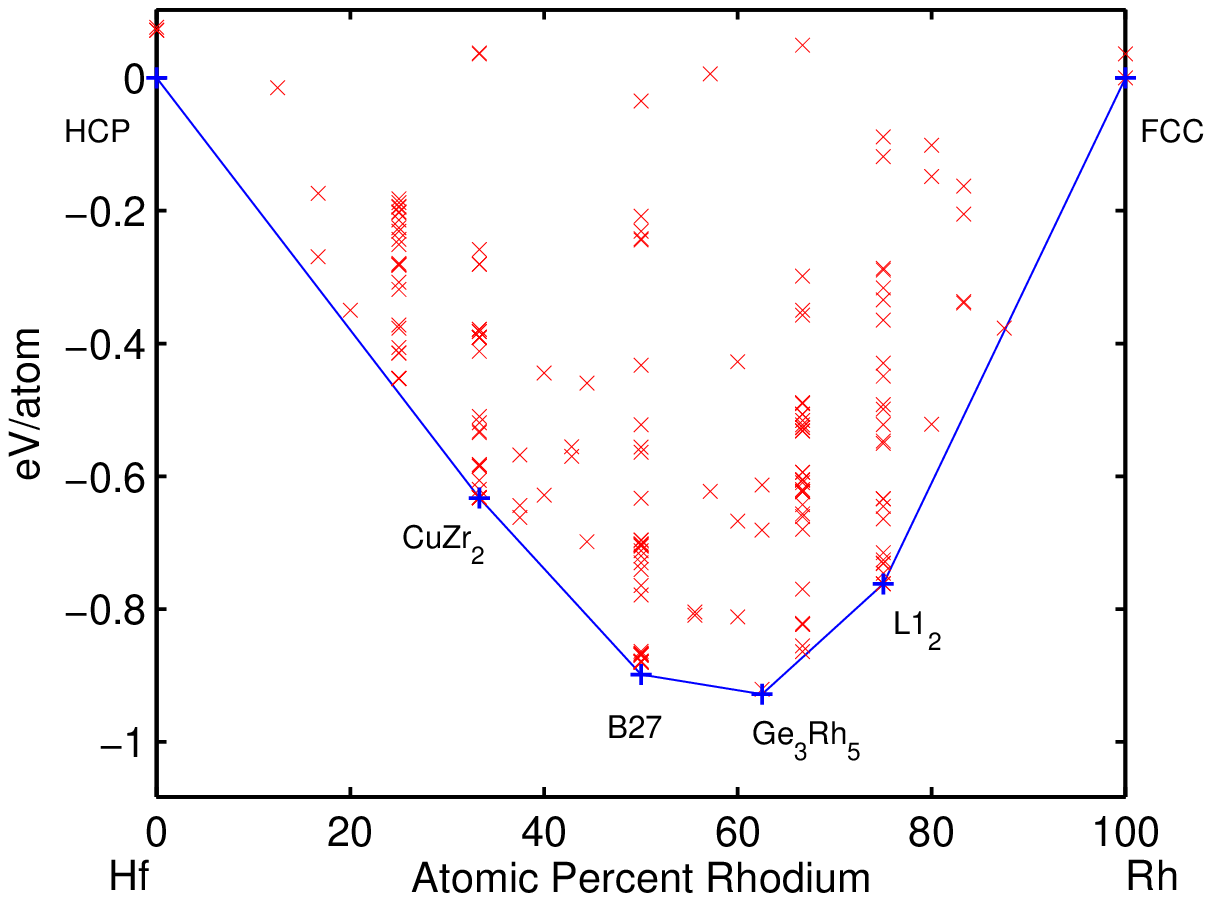}\hspace{\figsskip}
  \includegraphics[width=\figswidth]{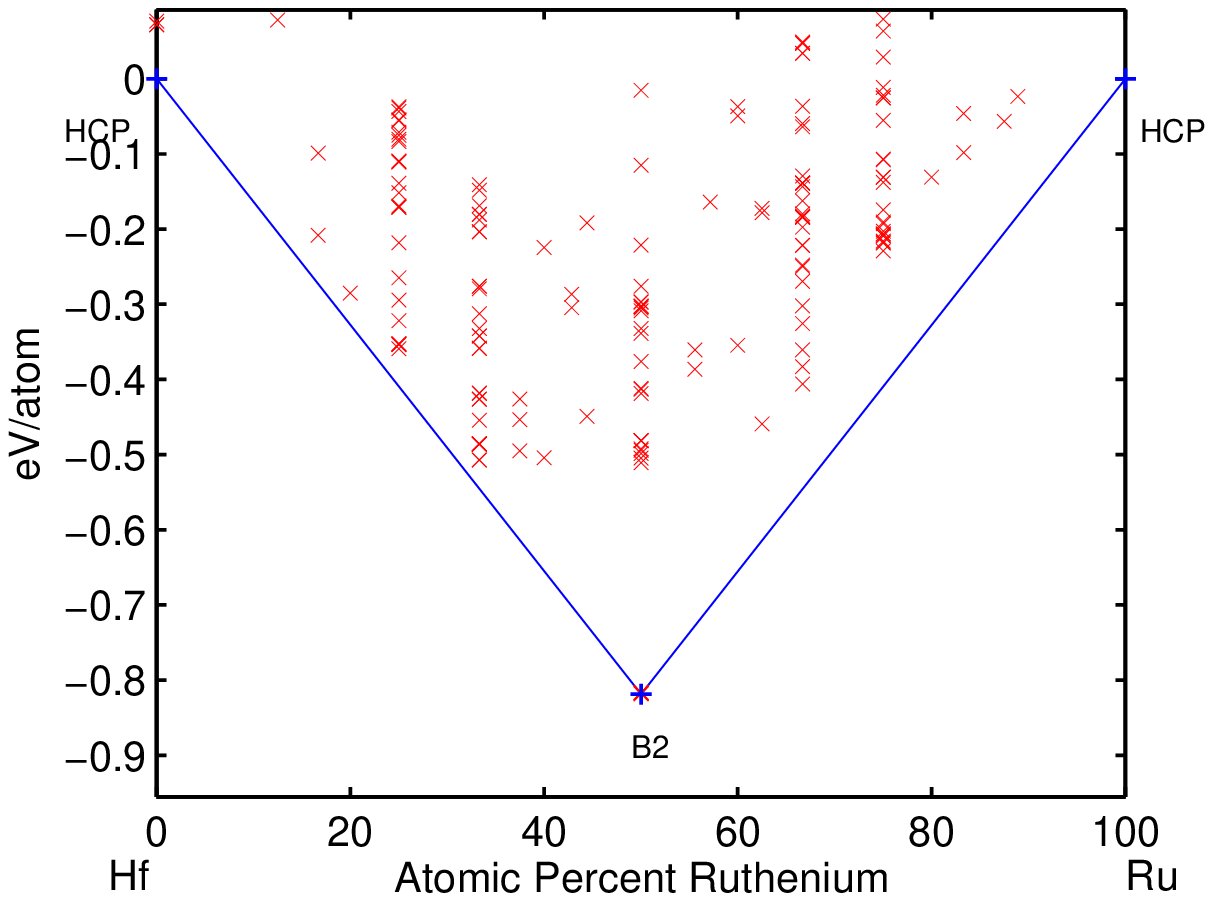}\hspace{\figsskip}
  \includegraphics[width=\figswidth]{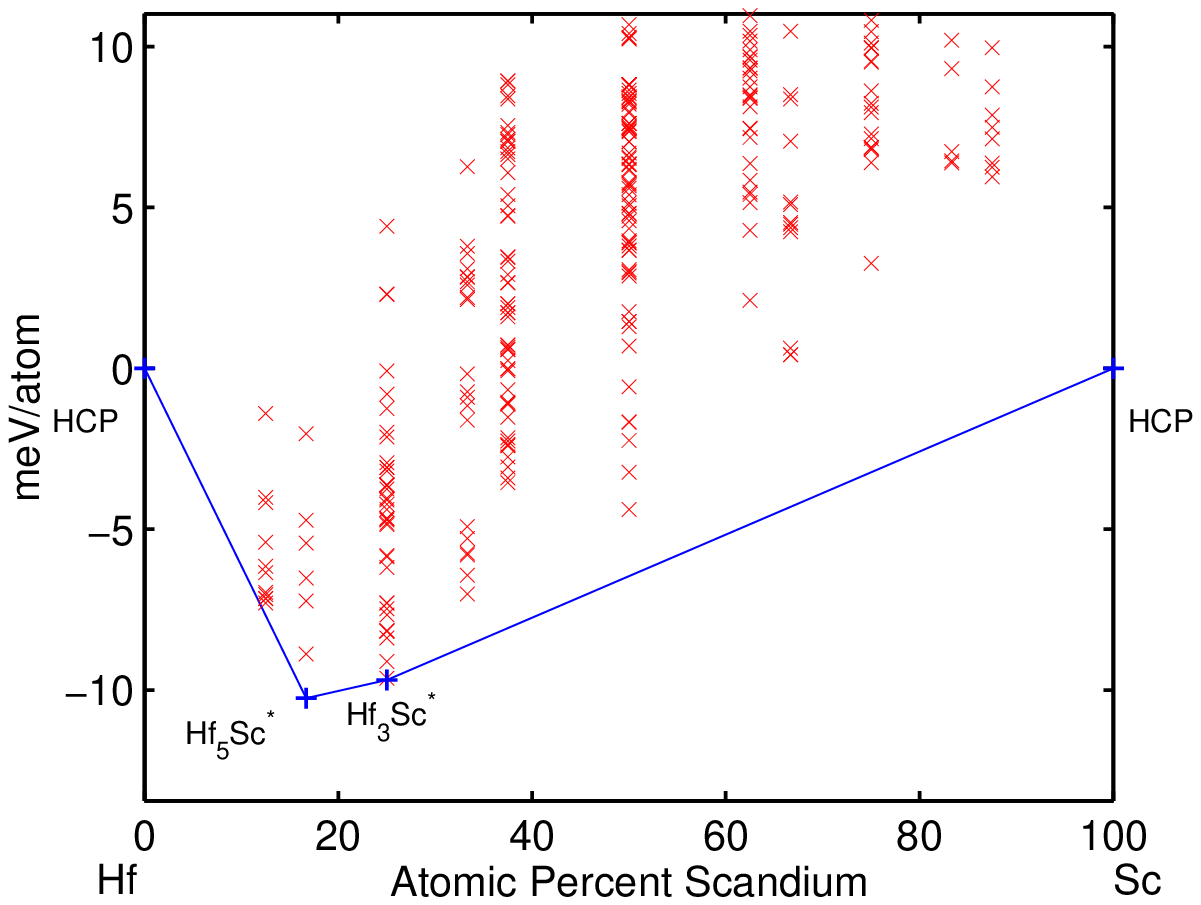}\hspace{\figsskip}
  \includegraphics[width=\figswidth]{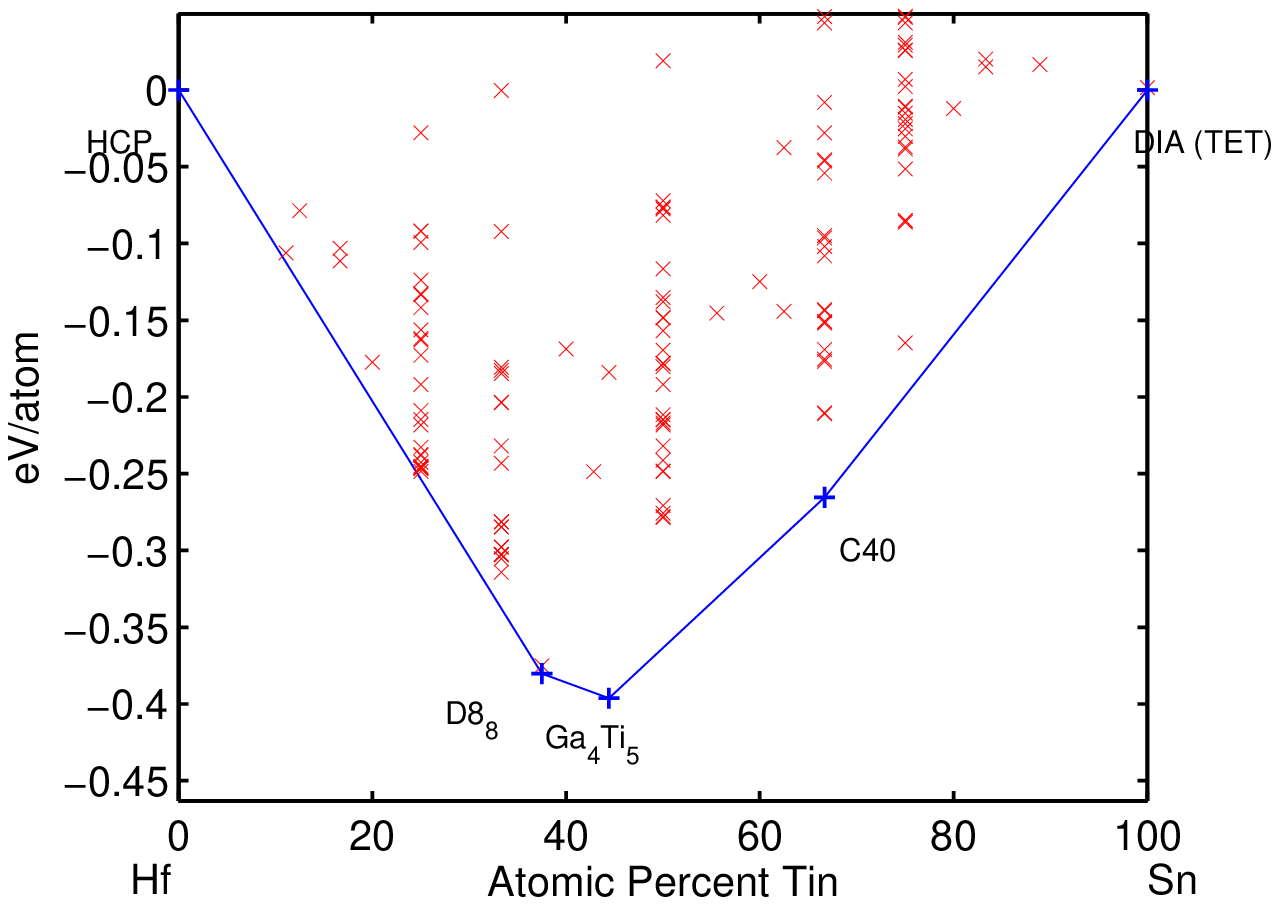}\hspace{\figsskip}
  \includegraphics[width=\figswidth]{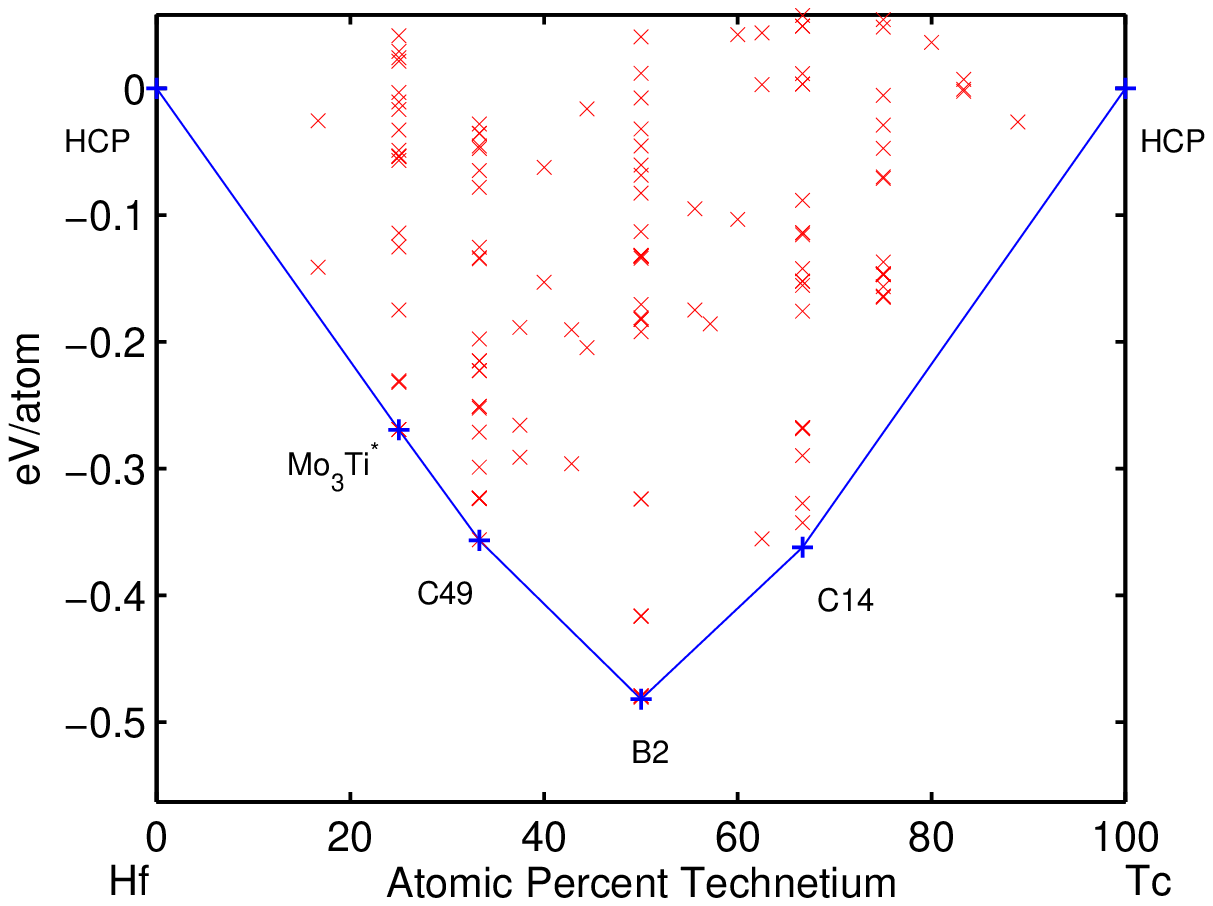}\hspace{\figsskip}
  \includegraphics[width=\figswidth]{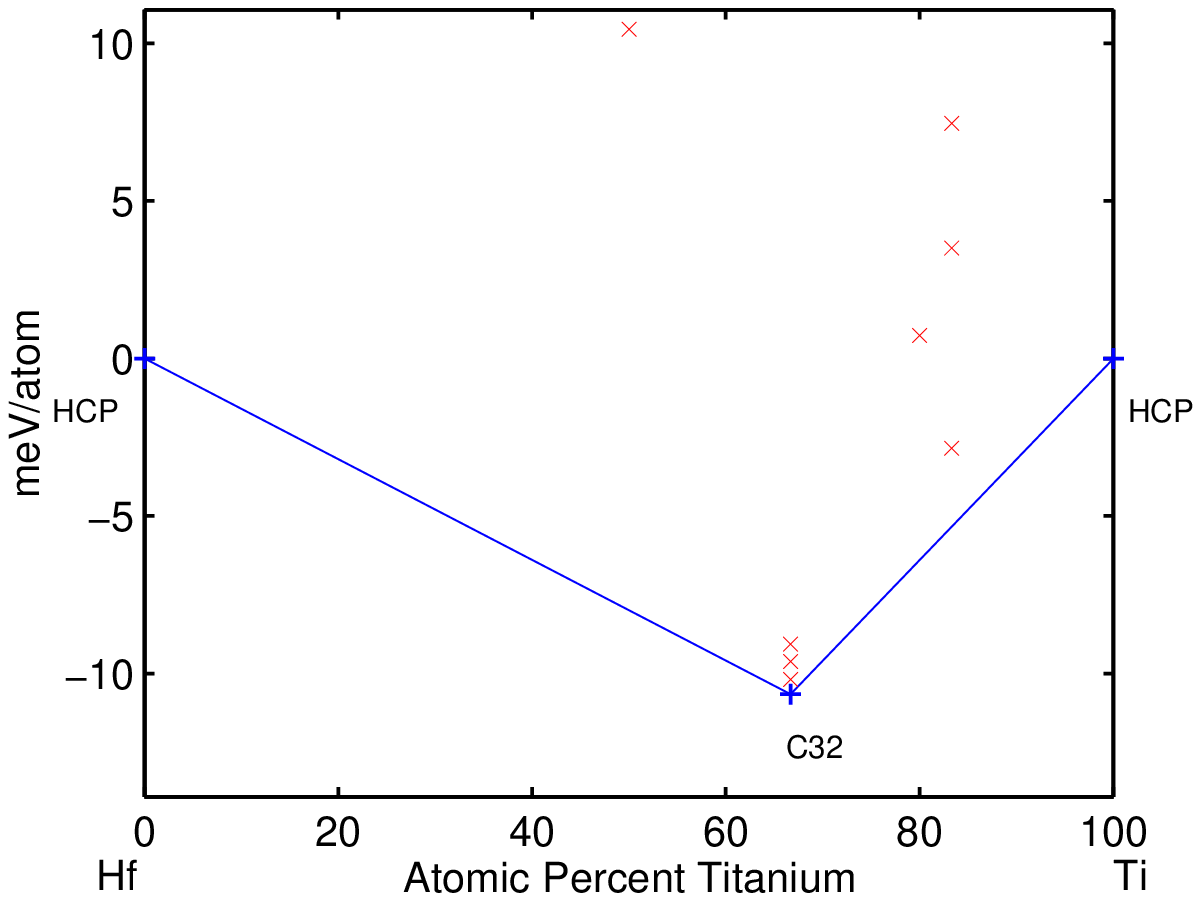}\hspace{\figsskip}
  \includegraphics[width=\figswidth]{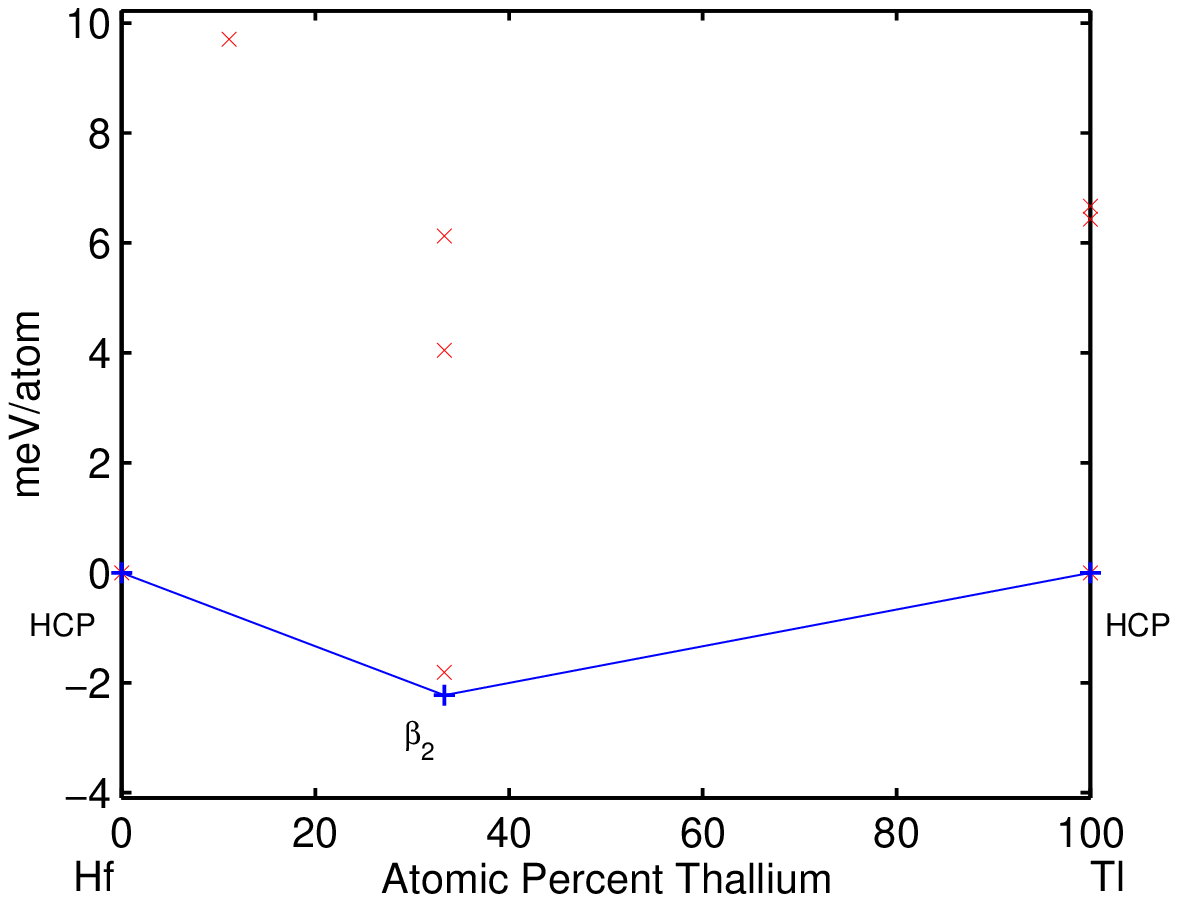}\hspace{\figsskip}
  \vnegspace
  \caption{{\small Formation enthalpies of
      Hf-Mn, Hf-Mo, Hf-Ni, 
      Hf-Os, Hf-Pb, Hf-Pd, 
      Hf-Pt, Hf-Re, Hf-Rh, 
      Hf-Ru, Hf-Sc, Hf-Sn, 
      Hf-Tc, Hf-Ti, and Hf-Tl 
      alloys.}}
  \vnegspace
  \label{fig_HfMn}\label{fig_HfMo}\label{fig_HfNi}
  \label{fig_HfOs}\label{fig_HfPb}\label{fig_HfPd}
  \label{fig_HfPt}\label{fig_HfRe}\label{fig_HfRh}
  \label{fig_HfRu}\label{fig_HfSc}\label{fig_HfSn}
  \label{fig_HfTc}\label{fig_HfTi}\label{fig_HfTl}
\end{figure}

}\end{widetext}

\clearpage

Following these findings we carried out a more extensive HT search for additional stable structures
in these six binary systems. This search included 137 fcc structures with up to six atoms per cell,
137 bcc structures with up to six atoms per cell and 333 hcp structures with up to eight atoms
per cell
(enumerated in Ref.\ \onlinecite{gus_enum}).  This search found lower energy structures in the Hf-Sc
system but failed to improve on the findings of the smaller set search in the other five systems.
This demonstrates the efficiency and reliability of the HT approach in finding phase-stability in
binary systems, using a search base of about 200 structures per system. More extensive searches, on
larger sets of derivative structures, seem to be useful only
in cases where both elements and their compounds share a common parent lattice.

In contrast to the preceding examples, our calculations predict phase-separation in the Hf-V
system, for which the compound HfV$_2$ has been reported at
high temperatures (1000-1550 $^\circ$C) \cite{Pauling,Massalski}.  Calculations of possible
structures of this compound show that the lowest energies, all positive, are obtained for C14, C36
and C15 in that order, whereas the structure reported in the literature is C15 \cite{Pauling}. This
is in agreement with the calculation in Ref. \onlinecite{Ormeci96} that also found C14 as the lowest
energy structure of these three Laves phases for HfV$_2$.  The compound HfV$_2$ reported in the
literature is therefore unstable at low temperatures.  The discrepancy between the computational
result of C14 as the minimum energy structure (Hf atoms on the sites of a hexagonal diamond
structure) and
the observed high temperature C15 phase (Hf on the sites of a diamond structure) may be due to
vibrational stabilization at high temperature.  In fact, C15 is only 24 meV/atom above C14, and
given that entropic differences between structures can be of the order of 0.1-1.0k$_B$ per atom
\cite{Axel_RMP}, very small energy differences between the experimentally observed structure and our
{\it ab initio} results could be reversed at elevated temperature. The phenomenon is common, e.g. in
Ref. \onlinecite{wolverton:prl_2001_AlCu} the vibrational entropy difference is shown to stabilize
the $\theta$-Al$_2$Cu (C16) phase over the competing Al$_2$Cu-$\theta'$ phase (distortion of
$\theta_c$-C1), which has the lowest energy and is, therefore, stable at low temperatures.

In agreement with experimental data, all the other transition metals, from columns VIB to IIB of the
periodic table, have stable compounds with hafnium at low temperatures.  The situation is especially
simple for the VIB metals, chromium, molybdenum and tungsten, for which the calculations confirm the
existence of a single compound Hf$M$$_2$ with a C15 structure.  Other Hf-Mo compounds for which {\it
  ab initio} calculations have been previously performed \cite{Kong05} are shown to be unstable.
The Hf-Hg system also exhibits a single compound, Hf$_2$Hg, where the calculations confirm that the
reported structures C11$_b$ \cite{Massalski} and CuZr$_2$ \cite{Pauling} are degenerate.

The calculations for the Fe-Hf system show the existence of a stable compound Fe$_2$Hf with a C15
structure, and the C14 structure just slightly less stable (14 meV/atom).  This is consistent with
the literature where the two structures have been reported \cite{Pauling} at temperatures above
600$^\circ$C.  However, in contrast to the available experimental data, the calculation shows that
the compounds FeHF (B2 structure) and Fe$_5$Hf (C15$_b$) are also stable whereas the FeHf$_2$
compound, reported with a NiTi$_2$ structure, is unstable at low temperatures.
Its formation energy is 35 meV/atom higher than the tie line FeHf(B2)$\leftrightarrow$Hf(A3) and thus
could decompose at temperatures lower than $100^\circ$C (Fig.  \ref{fig_FeHf}).  A similar
behavior is found in the HfMn system where the C14 structure of HfMn$_2$ is found to be stable, with
the C36 and C15 structures at slightly higher energies (4 and 5 meV/atom, respectively).  The
experimentally reported Hf$_2$Mn compound (NiTi$_2$ prototype), is again found to be unstable at low
temperatures, at 30meV/atom above the Hf(A3)$\leftrightarrow$HfMn$_2$(C14) tie line.

In the Hf-Ru and Hf-Os systems, a stable B2 structure of HfRu and HfOs is found
in agreement with the experiments \cite{Pauling}. The compounds Hf$_2$Os,
Hf$_{54}$Os$_{17}$ and HfOs$_2$ reported from experimental data are found to be unstable. No stable
structure was found for the HfRu$_2$ compound, whose existence was suspected in some experiments \cite{Pauling}.
In the Hf-Re system ref. \cite{Pauling} reports a Hf$_{21}$Re$_{25}$ compound (prototype Re$_{25}$Zr$_{21}$)
whereas ref. \cite{Massalski} reports a HfRe compound of unspecified prototype. Our calculations confirm
the existence of the Hf$_{21}$Re$_{25}$ compound but find no stable structure for HfRh. The
lowest HfRh structure, B2,
lies approximately 50meV/atom above the convex hull (Table \ref{table1} and Fig. \ref{fig_HfRe}).
For the Co-Hf system, our calculations confirm the existence of the stable compounds Co$_2$Hf, CoHf
and CoHf$_2$ at low temperatures, albeit with structures different from those reported in
experiments. The reported Ni$_7$Zr$_2$ structure of Co$_7$Hf$_2$ lies
above the convex hull (Table \ref{table1} and Fig. \ref{fig_CoHf}).

The Cu-Hf system is the only transition metal-hafnium system previously studied by {\it ab initio}
methods, also using the VASP software but with different potentials than in our study
(Vanderbilt-type ultrasoft pseudopotentials) \cite{Ghosh07}.  28 different compound structures have
been calculated, of which four, Cu$_5$Hf, Cu$_8$Hf$_3$, Cu$_{10}$Hf$_7$ and CuHf$_2$ were identified
as stable.  Our calculations also indicate that these four compounds are stable.  There is no
experimental information about the sructure of Cu$_5$Hf \cite{Pauling}.  We obtain the AuBe$_5$
structure in agreement with Ref.\ \cite{Ghosh07}.  The CuZr$_2$ structure, reported in experiments
for CuHf$_2$ \cite{Pauling}, and not calculated in ref. \cite{Ghosh07} is degenerate with the
reported C11$_b$ structure.  The ground state convex hull is asymmetric and skewed towards the Cu
side, in agreement with the general conclusions of Ref. \cite{Ghosh07} (Fig.\ \ref{fig_CuHf}).  The
Al-Hf system was studied using the same methodology \cite{Ghosh05}.  The convex hull in this case is
defined by four compounds Al$_3$Hf(D0$_{23}$), Al$_2$Hf(C14), Al$_3$Hf$_4$(Al$_3$Zr$_4$) and
AlHf$_3$(L1$_2$), with Al$_3$Hf$_2$(Al$_3$Zr$_2$) and AlHf(B33) slightly above the convex hull.
Our calculations reproduce these results and show in addition that the AlHf structure TlI, reported
in the experimental data but not studied in Ref. \cite{Ghosh05}, is degenerate with B33.  Other
compounds reported for this system \cite{Pauling} are found to be significantly above the convex
hull and should therefore be unstable at low temperatures (see Table \ref{table1} and Fig.\
\ref{fig_AlHf}).

Our HT study uncovers a few compounds on which no data is available in the experimental literature
\cite{Massalski,Pauling} in the binary systems of Hf with Au, Bi, In, Pd, Pt, Re, and Tc.  A few
reported compounds are found to lie above the convex hull of their respective
systems (Hf with Be, Ga, Ni and Sn) and are thus predicted to be unstable at low temperatures.
(Table \ref{table1})

In most cases where two structures are reported in the experimental literature, the calculations
show that they are indeed degenerate or the energy differences are small, 30 meV/atom or less
(square parentheses in Table \ref{table1}).  In only two cases these structures differ considerably.
The B2
structure of HfPt is 165meV/atom higher than the degenerate B33/TlI groundstate.  No phase diagram
is available for the Hf-Pt system, but the experimenatal data indicates that the B2 structure might
have been observed in a non-stoichiometric mixture \cite{Massalski}. The calculations seem to provide
an indirect confirmation of this observation, showing that B2 is not a stable structure of the
stoichiometric mixture.  The Th$_2$Ni$_{17}$ structure of Be$_{17}$Hf$_2$ is 485meV/atom higher than
the Th$_2$Zn$_{17}$ groundstate (sometimes also denoted as Be$_{17}$Nb$_2$ \cite{Pauling,Massalski}).
As discussed in Ref.\ \onlinecite{Massalski}, it has not been determined whether Be$_{17}$Hf$_2$
consists of both forms or just one of them. The large energy difference obtained in our calculations
indicates that one structure is much more likely than the other and the compound should consist
exclusively of the Th$_2$Zr$_{17}$ prototype.

In conclusion, a systematic and comprehensive {\it ab initio} study of phase stability is carried
out for the hafnium intermetallic systems. The total energies of about 200 intermetallic compounds have
been calculated for \nsystems hafnium-metal systems and some interesting deviations from published
experimental data were found. In particular, the calculations predict the existence of stable
compounds in six hafnium intermetallic systems previously believed to be phase-separating.  On the
Pettifor chemical scale, three of these are isolated phase-separating systems in a cluster of
compound-forming ones, and the discrepency is likely due to lack of experimental data.  Hence, these
predictions nicely complement the trend indicated by the empirical scale.  A few
new compounds are predicted in binary systems of Hf and other metals, and some compounds reported in
the literature are shown to be unstable at low temperatures.  

A detailed understanding of Hf alloys is crucial for a better realization of its potential as an
alloying agent in currently available applications and in developing new ones.  The picture of Hf
alloys that emerges from this study is quite different from that depicted by current experimental
data.  It should be emphasized that we consider the alloys to be in thermodynamical equilibrium,
which can be difficult to reach at low temperatures due to slow kinetics. At higher temperatures,
configurational disorder and vibrational entropic promotion might destabilize the predicted
compounds.  The theoretical predictions presented here should therefore motivate research for their
experimental validation and provide useful guidance to future studies of these alloys.

We thank Wahyu Setyawan and Mike Mehl for fruitful discussions.  
Research supported by ONR (Grants N00014-07-1-0878, N00014-07-1-1085, N00014-09-1-0921) and NSF (DMR-0639822, DMR-0650406). 
We thank the Teragrid Partnership (Texas advanced Computing Center, TACC, MCA-07S005) 
and the Fulton Supercomputing Laboratory at Brigham Young University for computational support.

\newpage

\begin{widetext}{

\begin{table}[thw]
  \caption{Geometry of new prototypes in our study.
    Positions are given as unrelaxed positions in the parent lattice.} \label{protos}
  \scriptsize
      {
    \hspace{-8mm}\begin{tabular}{||c|c|c|c|c|c|c||}\hline  
      System           & Hf$_3$Sc$^{\star}$ & Hf$_5$Sc$^{\star}$   & BiHf$_2^{\star}$      & Mo$_3$Ti$^{\star}$     &Hf$_5$Pb$^{\star}$  &  HfPd$_5^{\star}$  \\
      Superlattice     &  HCP A$_6$B$_2$    &  HCP A$_5$B$_1$      &  HCP A$_2$B$_4$       &  BCC AB$_3$            & FCC A$_5$B$_1$     &  none              \\ 
                       &                    &                      &                       & \cite{monster}        &                    &                    \\ \hline
      Lattice          &   Orthorhombic     &  Hexagonal           & Monoclinic               &  Orthorhombic       & Tetragonal         & Orthorhombic       \\ \hline 
      Space Group      &    Cmcm \#63       &  P$\bar{6}$2m \#189  & C2/m \#  12              &  Immm \#71          &  P4/mmm \#123      & Cmmm \#65          \\ \hline 
      Pearson symbol   &        oC16        &        hP6           & mC12                     &          oI8        &  tP6               & oC12               \\ \hline
      Primitive        &                    &                      &                          &                     &                    &                    \\ 
      vectors (cart.)  &                    &                      &                          &                     &                    &                    \\
      {\bf a$_1$}/a    & $(  2,0 ,0)$       &$(3/2,\sqrt{3}/2,0)$  &$(-1,0,0)$                & $( 3/2, 1/2,-1/2 )$ & $(1/2,1/2,0)$      & $(-2.01059,5.93236,0)$   \\
      {\bf a$_2$}/a    & $( 1,\sqrt{3},0)$  &$(3/2,3\sqrt{3}/2,0)$ &$(1/2,\sqrt{3}/2,\sqrt{8/3})$& $( 1/2, 3/2, 1/2 )$ & $(0,3,3)$       & $(-2.01059,-5.93236,0)$  \\
      {\bf a$_3$}/a    & $(0,0,\sqrt{8/3})$ &$(0,0,\sqrt{8/3})$    &$(0,\sqrt{3},-\sqrt{8/3})$& $(-1/2,-3/2, 1/2 )$ & $(1/2,5/2,3)$      & $(0,0,4.02718)$    \\  \hline 
      Atomic           &                    &                      &                          &                     &                    &                    \\ 
      positions (fract.) &                  &                      &                          &                     &                    &                    \\ 
      {\bf A$1$}       & $(  0, 0, 0)$      & $(   0, 0, 0)$       & $(0,0,0)$                &  $(   0, 0, 0 )$    & $(0,0,0)$          & $(0.02752,0.97247,0)$        \\
      {\bf A$2$}       & $(  0,1/2, 0)$     & $(0,1/3,0)$          & $(7/18,7/95/18)$         &           $-$       & $(0,1/6,0)$        & $-$          \\ 
      {\bf A$3$}       & $(1/2, 0, 0)$      & $(0,2/3,0)$          & $-$                      &           $-$       & $(0,1/3,0)$        & $-$          \\ 
      {\bf A$4$}       & $(1/6,1/6,1/2)$    & $(1/3,0,1/2)$        & $-$                      &           $-$       & $(0,1/2,0)$        & $-$          \\ 
      {\bf A$5$}       & $(1/6,2/3,1/2)$    & $(1/3,1/3,1/2)$      & $-$                      &           $-$       & $(0,2/3,0)$        & $-$          \\ 
      {\bf A$6$}       & $(2/3,1/6,1/2)$    &        $-$           & $-$                      &           $-$       & $-$                & $-$          \\ 
      {\bf B$1$}       & $(1/2,1/2, 0)$     & $(1/3,2/3,1/2)$      & $(1/3,2/3,2/3)$          &  $(1/4,3/4,1/2)$    & $(0,5/6,0)$        & $(0.19430,0.80569,0.5)$          \\
      {\bf B$2$}       & $(2/3,2/3,1/2)$    &        $-$           & $(2/3,1/3,1/3)$          &  $(1/2,1/2, 0)$     & $-$                & $(0.36452,0.63547,0)$            \\
      {\bf B$3$}       &        $-$         &        $-$           & $(13/18,4/9,17/18)$      &  $(3/4,1/4,1/2)$    & $-$                & $(0.52795,0.47204,0.5)$            \\ 
      {\bf B$4$}       &        $-$         &        $-$           & $(1/18,1/9,11/18)$       &           $-$       & $-$                & $(0.69061,0.30938,0)$          \\ 
      {\bf B$5$}       &        $-$         &        $-$           &        $-$               &           $-$       & $-$                & $(0.86173,0.13826,0.5)$          \\ \hline
  {\small AFLOW} label &  ``471''           & ``473''              & ``475''                  & ``81''              &  ``477''           & ``479''                       \\ 
     \cite{aflow}     &                    &                      &                          &                     &                    &                     \\ \hline
     \end{tabular}}
\end{table} 

}\end{widetext}


\begin{thebibliography}{51}
\expandafter\ifx\csname natexlab\endcsname\relax\def\natexlab#1{#1}\fi
\expandafter\ifx\csname bibnamefont\endcsname\relax
  \def\bibnamefont#1{#1}\fi
\expandafter\ifx\csname bibfnamefont\endcsname\relax
  \def\bibfnamefont#1{#1}\fi
\expandafter\ifx\csname citenamefont\endcsname\relax
  \def\citenamefont#1{#1}\fi
\expandafter\ifx\csname url\endcsname\relax
  \def\url#1{\texttt{#1}}\fi
\expandafter\ifx\csname urlprefix\endcsname\relax\def\urlprefix{URL }\fi
\providecommand{\bibinfo}[2]{#2}
\providecommand{\eprint}[2][]{\url{#2}}

\bibitem[{\citenamefont{J\'{o}hannesson
  et~al.}(2002)\citenamefont{J\'{o}hannesson, Bligaard, Ruban, Skriver,
  Jacobsen, and N{\o}rskov}}]{Johann02}
\bibinfo{author}{\bibfnamefont{G.~H.} \bibnamefont{J\'{o}hannesson}},
  \bibinfo{author}{\bibfnamefont{T.}~\bibnamefont{Bligaard}},
  \bibinfo{author}{\bibfnamefont{A.~V.} \bibnamefont{Ruban}},
  \bibinfo{author}{\bibfnamefont{H.~L.} \bibnamefont{Skriver}},
  \bibinfo{author}{\bibfnamefont{K.~W.} \bibnamefont{Jacobsen}},
  \bibnamefont{and} \bibinfo{author}{\bibfnamefont{J.~K.}
  \bibnamefont{N{\o}rskov}}, \bibinfo{journal}{Phys.\ Rev.\ Lett.}
  \textbf{\bibinfo{volume}{88}}, \bibinfo{pages}{255506}
  (\bibinfo{year}{2002}).

\bibitem[{\citenamefont{Stucke and Crespi}(2003)}]{Stucke03}
\bibinfo{author}{\bibfnamefont{D.~P.} \bibnamefont{Stucke}} \bibnamefont{and}
  \bibinfo{author}{\bibfnamefont{V.~H.} \bibnamefont{Crespi}},
  \bibinfo{journal}{Nano Lett.} \textbf{\bibinfo{volume}{3}},
  \bibinfo{pages}{1183} (\bibinfo{year}{2003}).

\bibitem[{\citenamefont{Curtarolo et~al.}(2003)\citenamefont{Curtarolo, Morgan,
  Persson, Rodgers, and Ceder}}]{curtarolo:prl_2003_datamining}
\bibinfo{author}{\bibfnamefont{S.}~\bibnamefont{Curtarolo}},
  \bibinfo{author}{\bibfnamefont{D.}~\bibnamefont{Morgan}},
  \bibinfo{author}{\bibfnamefont{K.}~\bibnamefont{Persson}},
  \bibinfo{author}{\bibfnamefont{J.}~\bibnamefont{Rodgers}}, \bibnamefont{and}
  \bibinfo{author}{\bibfnamefont{G.}~\bibnamefont{Ceder}},
  \bibinfo{journal}{Phys.\ Rev.\ Lett.} \textbf{\bibinfo{volume}{91}},
  \bibinfo{pages}{135503} (\bibinfo{year}{2003}).

\bibitem[{\citenamefont{Curtarolo et~al.}(2005)\citenamefont{Curtarolo, Morgan,
  and Ceder}}]{monster}
\bibinfo{author}{\bibfnamefont{S.}~\bibnamefont{Curtarolo}},
  \bibinfo{author}{\bibfnamefont{D.}~\bibnamefont{Morgan}}, \bibnamefont{and}
  \bibinfo{author}{\bibfnamefont{G.}~\bibnamefont{Ceder}},
  \bibinfo{journal}{Calphad} \textbf{\bibinfo{volume}{29}},
  \bibinfo{pages}{163} (\bibinfo{year}{2005}).

\bibitem[{\citenamefont{Fischer et~al.}(2006)\citenamefont{Fischer, Tibbetts,
  Morgan, and Ceder}}]{Fischer06}
\bibinfo{author}{\bibfnamefont{C.~C.} \bibnamefont{Fischer}},
  \bibinfo{author}{\bibfnamefont{K.~J.} \bibnamefont{Tibbetts}},
  \bibinfo{author}{\bibfnamefont{D.}~\bibnamefont{Morgan}}, \bibnamefont{and}
  \bibinfo{author}{\bibfnamefont{G.}~\bibnamefont{Ceder}},
  \bibinfo{journal}{Nature Materials} \textbf{\bibinfo{volume}{5}},
  \bibinfo{pages}{641} (\bibinfo{year}{2006}).

\bibitem[{\citenamefont{Lewis et~al.}(2007)\citenamefont{Lewis, Sachtler, Low,
  Lesch, Faheem, Dosek, Knight, Halloran, Jensen, Yang et~al.}}]{Lewis2007355}
\bibinfo{author}{\bibfnamefont{G.~J.} \bibnamefont{Lewis}},
  \bibinfo{author}{\bibfnamefont{J.~W.~A.} \bibnamefont{Sachtler}},
  \bibinfo{author}{\bibfnamefont{J.~J.} \bibnamefont{Low}},
  \bibinfo{author}{\bibfnamefont{D.~A.} \bibnamefont{Lesch}},
  \bibinfo{author}{\bibfnamefont{S.~A.} \bibnamefont{Faheem}},
  \bibinfo{author}{\bibfnamefont{P.~M.} \bibnamefont{Dosek}},
  \bibinfo{author}{\bibfnamefont{L.~M.} \bibnamefont{Knight}},
  \bibinfo{author}{\bibfnamefont{L.}~\bibnamefont{Halloran}},
  \bibinfo{author}{\bibfnamefont{C.~M.} \bibnamefont{Jensen}},
  \bibinfo{author}{\bibfnamefont{J.}~\bibnamefont{Yang}}, \bibnamefont{et~al.},
  \bibinfo{journal}{J. Alloys Compound.} \textbf{\bibinfo{volume}{446-447}},
  \bibinfo{pages}{355 } (\bibinfo{year}{2007}).

\bibitem[{\citenamefont{Ozolins et~al.}(2008)\citenamefont{Ozolins, Majzoub,
  and Wolverton}}]{ozolins:135501}
\bibinfo{author}{\bibfnamefont{V.}~\bibnamefont{Ozolins}},
  \bibinfo{author}{\bibfnamefont{E.~H.} \bibnamefont{Majzoub}},
  \bibnamefont{and}
  \bibinfo{author}{\bibfnamefont{C.}~\bibnamefont{Wolverton}},
  \bibinfo{journal}{Phys.\ Rev.\ Lett.} \textbf{\bibinfo{volume}{100}},
  \bibinfo{eid}{135501} (\bibinfo{year}{2008}).

\bibitem[{\citenamefont{C.Ortiz et~al.}(2009)\citenamefont{C.Ortiz, Eriksson,
  and Klintenberg}}]{Ortiz09}
\bibinfo{author}{\bibnamefont{C.Ortiz}},
  \bibinfo{author}{\bibfnamefont{O.}~\bibnamefont{Eriksson}}, \bibnamefont{and}
  \bibinfo{author}{\bibfnamefont{M.}~\bibnamefont{Klintenberg}},
  \bibinfo{journal}{Comput. Mater. Sci.} \textbf{\bibinfo{volume}{44}},
  \bibinfo{pages}{1042} (\bibinfo{year}{2009}).

\bibitem[{\citenamefont{Cramer and Covino}(2005)}]{ASMHandbook}
\bibinfo{editor}{\bibfnamefont{S.~D.} \bibnamefont{Cramer}} \bibnamefont{and}
  \bibinfo{editor}{\bibfnamefont{B.~S.} \bibnamefont{Covino}}, eds.,
  \emph{\bibinfo{title}{ASM Handbook}}, vol. \bibinfo{volume}{13B}
  (\bibinfo{publisher}{ASM International}, \bibinfo{address}{Metals Park,
  Ohio}, \bibinfo{year}{2005}), \bibinfo{edition}{10th} ed.

\bibitem[{\citenamefont{Wallenius and Westlen}(2008)}]{Wallenius08}
\bibinfo{author}{\bibfnamefont{J.}~\bibnamefont{Wallenius}} \bibnamefont{and}
  \bibinfo{author}{\bibfnamefont{D.}~\bibnamefont{Westlen}},
  \bibinfo{journal}{Ann. Nucl. Energy} \textbf{\bibinfo{volume}{35}},
  \bibinfo{pages}{60} (\bibinfo{year}{2008}).

\bibitem[{\citenamefont{Nielsen}(1989)}]{Nielsen_Hf}
\bibinfo{author}{\bibfnamefont{R.~H.} \bibnamefont{Nielsen}}, in
  \emph{\bibinfo{booktitle}{Ullman's Encyclopedia of Industrial Chemistry}}
  (\bibinfo{publisher}{VCH Verlagsgesellschaft mbH}, \bibinfo{year}{1989}),
  vol. \bibinfo{volume}{A12}, pp. \bibinfo{pages}{559--569}.

\bibitem[{\citenamefont{Davidson}(Sept.21, 1999)}]{patent_5954724}
\bibinfo{author}{\bibfnamefont{J.~A.} \bibnamefont{Davidson}},
  \emph{\bibinfo{title}{Titanium molybdenum hafnium alloys for medical implants
  and devices}}, \bibinfo{howpublished}{US Patent 5954724}
  (\bibinfo{year}{Sept.21, 1999}).

\bibitem[{\citenamefont{Meng et~al.}(2009)\citenamefont{Meng, Fu, Cai, Li, and
  Zhao}}]{Meng09}
\bibinfo{author}{\bibfnamefont{X.~L.} \bibnamefont{Meng}},
  \bibinfo{author}{\bibfnamefont{Y.~D.} \bibnamefont{Fu}},
  \bibinfo{author}{\bibfnamefont{W.}~\bibnamefont{Cai}},
  \bibinfo{author}{\bibfnamefont{Q.~F.} \bibnamefont{Li}}, \bibnamefont{and}
  \bibinfo{author}{\bibfnamefont{L.~C.} \bibnamefont{Zhao}},
  \bibinfo{journal}{Phil. Mag. Lett.} \textbf{\bibinfo{volume}{89}},
  \bibinfo{pages}{431} (\bibinfo{year}{2009}).

\bibitem[{\citenamefont{Fernandes}(World Intelectual Property Organization,
  2001)}]{Patent_012868}
\bibinfo{author}{\bibfnamefont{M.~T.} \bibnamefont{Fernandes}},
  \emph{\bibinfo{title}{Aluminum-magnesium-scandium alloys with hafnium}},
  \bibinfo{howpublished}{WO/2001/012868} (\bibinfo{year}{World Intelectual
  Property Organization, 2001}).

\bibitem[{\citenamefont{Baudry et~al.}(1992)\citenamefont{Baudry, Boyer,
  Ferreira, Harris, Miraglia, and Pontonnier}}]{Baudry92}
\bibinfo{author}{\bibfnamefont{A.}~\bibnamefont{Baudry}},
  \bibinfo{author}{\bibfnamefont{P.}~\bibnamefont{Boyer}},
  \bibinfo{author}{\bibfnamefont{L.~P.} \bibnamefont{Ferreira}},
  \bibinfo{author}{\bibfnamefont{S.~W.} \bibnamefont{Harris}},
  \bibinfo{author}{\bibfnamefont{S.}~\bibnamefont{Miraglia}}, \bibnamefont{and}
  \bibinfo{author}{\bibfnamefont{L.}~\bibnamefont{Pontonnier}},
  \bibinfo{journal}{J.\ Phys.:\ Conden. Matt.} \textbf{\bibinfo{volume}{4}},
  \bibinfo{pages}{5025} (\bibinfo{year}{1992}).

\bibitem[{\citenamefont{Callegari et~al.}(2004)\citenamefont{Callegari,
  Jamison, Carrier, Zafar, Gusev, Narayanan, D'Emic, Lacey, Feely, Jammy
  et~al.}}]{Callegari04}
\bibinfo{author}{\bibfnamefont{A.}~\bibnamefont{Callegari}},
  \bibinfo{author}{\bibfnamefont{P.}~\bibnamefont{Jamison}},
  \bibinfo{author}{\bibfnamefont{E.}~\bibnamefont{Carrier}},
  \bibinfo{author}{\bibfnamefont{S.}~\bibnamefont{Zafar}},
  \bibinfo{author}{\bibfnamefont{E.}~\bibnamefont{Gusev}},
  \bibinfo{author}{\bibfnamefont{V.}~\bibnamefont{Narayanan}},
  \bibinfo{author}{\bibfnamefont{C.}~\bibnamefont{D'Emic}},
  \bibinfo{author}{\bibfnamefont{D.}~\bibnamefont{Lacey}},
  \bibinfo{author}{\bibfnamefont{F.~M.} \bibnamefont{Feely}},
  \bibinfo{author}{\bibfnamefont{R.}~\bibnamefont{Jammy}},
  \bibnamefont{et~al.}, in \emph{\bibinfo{booktitle}{Electron Devices Meeting,
  2004. IEDM Technical Digest. IEEE International}} (\bibinfo{publisher}{IEEE},
  \bibinfo{address}{Piscataway, NJ}, \bibinfo{year}{2004}), pp.
  \bibinfo{pages}{825--828}.

\bibitem[{\citenamefont{Pignedoli et~al.}(2007)\citenamefont{Pignedoli,
  Curioni, and Andreoni}}]{Pignedoli07}
\bibinfo{author}{\bibfnamefont{C.~A.} \bibnamefont{Pignedoli}},
  \bibinfo{author}{\bibfnamefont{A.}~\bibnamefont{Curioni}}, \bibnamefont{and}
  \bibinfo{author}{\bibfnamefont{W.}~\bibnamefont{Andreoni}},
  \bibinfo{journal}{Phys.\ Rev.\ Lett.} \textbf{\bibinfo{volume}{98}},
  \bibinfo{pages}{037602} (\bibinfo{year}{2007}).

\bibitem[{\citenamefont{Zhang and E.Evetts}(1994)}]{Zhang94}
\bibinfo{author}{\bibfnamefont{J.~L.} \bibnamefont{Zhang}} \bibnamefont{and}
  \bibinfo{author}{\bibfnamefont{J.}~\bibnamefont{E.Evetts}},
  \bibinfo{journal}{J. Mater. Sci.} \textbf{\bibinfo{volume}{29}},
  \bibinfo{pages}{778} (\bibinfo{year}{1994}).

\bibitem[{\citenamefont{Ghosh et~al.}(2002)\citenamefont{Ghosh, van~de Walle,
  Asta, and Olson}}]{Ghosh02}
\bibinfo{author}{\bibfnamefont{G.}~\bibnamefont{Ghosh}},
  \bibinfo{author}{\bibfnamefont{A.}~\bibnamefont{van~de Walle}},
  \bibinfo{author}{\bibfnamefont{M.}~\bibnamefont{Asta}}, \bibnamefont{and}
  \bibinfo{author}{\bibfnamefont{G.~B.} \bibnamefont{Olson}},
  \bibinfo{journal}{Calphad} \textbf{\bibinfo{volume}{26}},
  \bibinfo{pages}{491} (\bibinfo{year}{2002}).

\bibitem[{\citenamefont{Kong and Liu}(2005)}]{Kong05}
\bibinfo{author}{\bibfnamefont{L.~T.} \bibnamefont{Kong}} \bibnamefont{and}
  \bibinfo{author}{\bibfnamefont{B.~X.} \bibnamefont{Liu}},
  \bibinfo{journal}{J. Phys. Soc. Japan} \textbf{\bibinfo{volume}{74}},
  \bibinfo{pages}{1766} (\bibinfo{year}{2005}).

\bibitem[{\citenamefont{Chen et~al.}(2005)\citenamefont{Chen, Wolf, Podloucky,
  and Rogl}}]{XQChen05}
\bibinfo{author}{\bibfnamefont{X.~Q.} \bibnamefont{Chen}},
  \bibinfo{author}{\bibfnamefont{W.}~\bibnamefont{Wolf}},
  \bibinfo{author}{\bibfnamefont{R.}~\bibnamefont{Podloucky}},
  \bibnamefont{and} \bibinfo{author}{\bibfnamefont{P.}~\bibnamefont{Rogl}},
  \bibinfo{journal}{Phys.\ Rev.\ B} \textbf{\bibinfo{volume}{71}},
  \bibinfo{pages}{174101} (\bibinfo{year}{2005}).

\bibitem[{\citenamefont{Belosevi\'c-Cavor
  et~al.}(2006)\citenamefont{Belosevi\'c-Cavor, Koteski, Novakovic, Concas,
  Congiu, and Spano}}]{JBC06}
\bibinfo{author}{\bibfnamefont{J.}~\bibnamefont{Belosevi\'c-Cavor}},
  \bibinfo{author}{\bibfnamefont{V.}~\bibnamefont{Koteski}},
  \bibinfo{author}{\bibfnamefont{N.}~\bibnamefont{Novakovic}},
  \bibinfo{author}{\bibfnamefont{G.}~\bibnamefont{Concas}},
  \bibinfo{author}{\bibfnamefont{F.}~\bibnamefont{Congiu}}, \bibnamefont{and}
  \bibinfo{author}{\bibfnamefont{G.}~\bibnamefont{Spano}},
  \bibinfo{journal}{Eur. Phys. J. B} \textbf{\bibinfo{volume}{50}},
  \bibinfo{pages}{425} (\bibinfo{year}{2006}).

\bibitem[{\citenamefont{Zhang et~al.}(2008{\natexlab{a}})\citenamefont{Zhang,
  Zhang, Wang, Li, Dong, Xing, Guo, and Li}}]{CZhang08}
\bibinfo{author}{\bibfnamefont{C.}~\bibnamefont{Zhang}},
  \bibinfo{author}{\bibfnamefont{Z.}~\bibnamefont{Zhang}},
  \bibinfo{author}{\bibfnamefont{S.}~\bibnamefont{Wang}},
  \bibinfo{author}{\bibfnamefont{H.}~\bibnamefont{Li}},
  \bibinfo{author}{\bibfnamefont{J.}~\bibnamefont{Dong}},
  \bibinfo{author}{\bibfnamefont{N.}~\bibnamefont{Xing}},
  \bibinfo{author}{\bibfnamefont{Y.}~\bibnamefont{Guo}}, \bibnamefont{and}
  \bibinfo{author}{\bibfnamefont{W.}~\bibnamefont{Li}}, \bibinfo{journal}{J.
  Alloys Compound.} \textbf{\bibinfo{volume}{448}}, \bibinfo{pages}{53}
  (\bibinfo{year}{2008}{\natexlab{a}}).

\bibitem[{\citenamefont{Ormeci et~al.}(1996)\citenamefont{Ormeci, Chu,
  m.~Wills, Mitchell, Albers, Thoma, and Chen}}]{Ormeci96}
\bibinfo{author}{\bibfnamefont{A.}~\bibnamefont{Ormeci}},
  \bibinfo{author}{\bibfnamefont{F.}~\bibnamefont{Chu}},
  \bibinfo{author}{\bibfnamefont{J.}~\bibnamefont{m.~Wills}},
  \bibinfo{author}{\bibfnamefont{T.~E.} \bibnamefont{Mitchell}},
  \bibinfo{author}{\bibfnamefont{R.~C.} \bibnamefont{Albers}},
  \bibinfo{author}{\bibfnamefont{D.~J.} \bibnamefont{Thoma}}, \bibnamefont{and}
  \bibinfo{author}{\bibfnamefont{S.~P.} \bibnamefont{Chen}},
  \bibinfo{journal}{Phys.\ Rev.\ B} \textbf{\bibinfo{volume}{54}},
  \bibinfo{pages}{12753} (\bibinfo{year}{1996}).

\bibitem[{\citenamefont{Chen et~al.}(2004)\citenamefont{Chen, Wolf, Podloucky,
  and Rogl}}]{XQChen04}
\bibinfo{author}{\bibfnamefont{X.~Q.} \bibnamefont{Chen}},
  \bibinfo{author}{\bibfnamefont{W.}~\bibnamefont{Wolf}},
  \bibinfo{author}{\bibfnamefont{R.}~\bibnamefont{Podloucky}},
  \bibnamefont{and} \bibinfo{author}{\bibfnamefont{P.}~\bibnamefont{Rogl}},
  \bibinfo{journal}{J. Alloys Compound.} \textbf{\bibinfo{volume}{383}},
  \bibinfo{pages}{228} (\bibinfo{year}{2004}).

\bibitem[{\citenamefont{Lali\'c et~al.}(1999)\citenamefont{Lali\'c, Popovi\'c,
  and R.Vukajlovi\'c}}]{Lalic99}
\bibinfo{author}{\bibfnamefont{M.~V.} \bibnamefont{Lali\'c}},
  \bibinfo{author}{\bibfnamefont{Z.~S.} \bibnamefont{Popovi\'c}},
  \bibnamefont{and}
  \bibinfo{author}{\bibfnamefont{F.}~\bibnamefont{R.Vukajlovi\'c}},
  \bibinfo{journal}{J.\ Phys.:\ Conden. Matt.} \textbf{\bibinfo{volume}{11}},
  \bibinfo{pages}{2513} (\bibinfo{year}{1999}).

\bibitem[{\citenamefont{Belosevi\'c-Cavor
  et~al.}(2005)\citenamefont{Belosevi\'c-Cavor, Koteski, Concas, Cekic,
  Novakovic, and Spano}}]{JBC05}
\bibinfo{author}{\bibfnamefont{J.}~\bibnamefont{Belosevi\'c-Cavor}},
  \bibinfo{author}{\bibfnamefont{V.}~\bibnamefont{Koteski}},
  \bibinfo{author}{\bibfnamefont{G.}~\bibnamefont{Concas}},
  \bibinfo{author}{\bibfnamefont{B.}~\bibnamefont{Cekic}},
  \bibinfo{author}{\bibfnamefont{N.}~\bibnamefont{Novakovic}},
  \bibnamefont{and} \bibinfo{author}{\bibfnamefont{G.}~\bibnamefont{Spano}},
  \bibinfo{journal}{J. Phys. Chem. Solids} \textbf{\bibinfo{volume}{66}},
  \bibinfo{pages}{1815} (\bibinfo{year}{2005}).

\bibitem[{\citenamefont{Ceki\'c et~al.}(2004)\citenamefont{Ceki\'c, Ivanovi\'c,
  Koteski, Koicki, and Manasijevi\'c}}]{Cekic04}
\bibinfo{author}{\bibfnamefont{B.}~\bibnamefont{Ceki\'c}},
  \bibinfo{author}{\bibfnamefont{N.}~\bibnamefont{Ivanovi\'c}},
  \bibinfo{author}{\bibfnamefont{V.}~\bibnamefont{Koteski}},
  \bibinfo{author}{\bibfnamefont{S.}~\bibnamefont{Koicki}}, \bibnamefont{and}
  \bibinfo{author}{\bibfnamefont{M.}~\bibnamefont{Manasijevi\'c}},
  \bibinfo{journal}{J.\ Phys.:\ Conden. Matt.} \textbf{\bibinfo{volume}{16}},
  \bibinfo{pages}{3015} (\bibinfo{year}{2004}).

\bibitem[{\citenamefont{Hao et~al.}(2007)\citenamefont{Hao, Wu, Xu, Zhou, Liu,
  and Meng}}]{Hao07}
\bibinfo{author}{\bibfnamefont{X.}~\bibnamefont{Hao}},
  \bibinfo{author}{\bibfnamefont{Z.}~\bibnamefont{Wu}},
  \bibinfo{author}{\bibfnamefont{Y.}~\bibnamefont{Xu}},
  \bibinfo{author}{\bibfnamefont{D.}~\bibnamefont{Zhou}},
  \bibinfo{author}{\bibfnamefont{X.}~\bibnamefont{Liu}}, \bibnamefont{and}
  \bibinfo{author}{\bibfnamefont{J.}~\bibnamefont{Meng}}, \bibinfo{journal}{J.\
  Phys.:\ Conden. Matt.} \textbf{\bibinfo{volume}{19}}, \bibinfo{pages}{196212}
  (\bibinfo{year}{2007}).

\bibitem[{\citenamefont{Shein and L.Ivanovskii}(2008)}]{Shein08}
\bibinfo{author}{\bibfnamefont{I.~R.} \bibnamefont{Shein}} \bibnamefont{and}
  \bibinfo{author}{\bibfnamefont{A.}~\bibnamefont{L.Ivanovskii}},
  \bibinfo{journal}{J.\ Phys.:\ Conden. Matt.} \textbf{\bibinfo{volume}{20}},
  \bibinfo{pages}{415218} (\bibinfo{year}{2008}).

\bibitem[{\citenamefont{Zhang et~al.}(2008{\natexlab{b}})\citenamefont{Zhang,
  Luo, Han, Li, and Han}}]{XZhang08}
\bibinfo{author}{\bibfnamefont{X.}~\bibnamefont{Zhang}},
  \bibinfo{author}{\bibfnamefont{X.}~\bibnamefont{Luo}},
  \bibinfo{author}{\bibfnamefont{J.}~\bibnamefont{Han}},
  \bibinfo{author}{\bibfnamefont{J.}~\bibnamefont{Li}}, \bibnamefont{and}
  \bibinfo{author}{\bibfnamefont{W.}~\bibnamefont{Han}},
  \bibinfo{journal}{Comp. Mat. Sci.} \textbf{\bibinfo{volume}{44}},
  \bibinfo{pages}{411} (\bibinfo{year}{2008}{\natexlab{b}}).

\bibitem[{\citenamefont{Lali\'c et~al.}(1998)\citenamefont{Lali\'c, Ceki\'c,
  Popovi\'c, and R.Vukajlovi\'c}}]{Lalic98}
\bibinfo{author}{\bibfnamefont{M.~V.} \bibnamefont{Lali\'c}},
  \bibinfo{author}{\bibfnamefont{B.}~\bibnamefont{Ceki\'c}},
  \bibinfo{author}{\bibfnamefont{Z.~S.} \bibnamefont{Popovi\'c}},
  \bibnamefont{and}
  \bibinfo{author}{\bibfnamefont{F.}~\bibnamefont{R.Vukajlovi\'c}},
  \bibinfo{journal}{J.\ Phys.:\ Conden. Matt.} \textbf{\bibinfo{volume}{10}},
  \bibinfo{pages}{6285} (\bibinfo{year}{1998}).

\bibitem[{\citenamefont{Oh et~al.}(2008)\citenamefont{Oh, Kima, Park, Weeb, and
  Lee}}]{Oh08}
\bibinfo{author}{\bibfnamefont{M.~W.} \bibnamefont{Oh}},
  \bibinfo{author}{\bibfnamefont{B.~S.} \bibnamefont{Kima}},
  \bibinfo{author}{\bibfnamefont{S.~D.} \bibnamefont{Park}},
  \bibinfo{author}{\bibfnamefont{D.~M.} \bibnamefont{Weeb}}, \bibnamefont{and}
  \bibinfo{author}{\bibfnamefont{H.~W.} \bibnamefont{Lee}},
  \bibinfo{journal}{Solid State Comm.} \textbf{\bibinfo{volume}{146}},
  \bibinfo{pages}{454} (\bibinfo{year}{2008}).

\bibitem[{\citenamefont{Chen and Podloucky}(2006)}]{XQChen06}
\bibinfo{author}{\bibfnamefont{X.~Q.} \bibnamefont{Chen}} \bibnamefont{and}
  \bibinfo{author}{\bibfnamefont{R.}~\bibnamefont{Podloucky}},
  \bibinfo{journal}{Calphad} \textbf{\bibinfo{volume}{30}},
  \bibinfo{pages}{266} (\bibinfo{year}{2006}).

\bibitem[{\citenamefont{Ghosh and Asta}(2005)}]{Ghosh05}
\bibinfo{author}{\bibfnamefont{G.}~\bibnamefont{Ghosh}} \bibnamefont{and}
  \bibinfo{author}{\bibfnamefont{M.}~\bibnamefont{Asta}},
  \bibinfo{journal}{Acta Mat.} \textbf{\bibinfo{volume}{53}},
  \bibinfo{pages}{3225} (\bibinfo{year}{2005}).

\bibitem[{\citenamefont{Ghosh et~al.}(2008)\citenamefont{Ghosh, van~de Walle,
  and Asta}}]{Ghosh08}
\bibinfo{author}{\bibfnamefont{G.}~\bibnamefont{Ghosh}},
  \bibinfo{author}{\bibfnamefont{A.}~\bibnamefont{van~de Walle}},
  \bibnamefont{and} \bibinfo{author}{\bibfnamefont{M.}~\bibnamefont{Asta}},
  \bibinfo{journal}{Acta Mat.} \textbf{\bibinfo{volume}{56}},
  \bibinfo{pages}{3202} (\bibinfo{year}{2008}).

\bibitem[{\citenamefont{Ghosh}(2007)}]{Ghosh07}
\bibinfo{author}{\bibfnamefont{G.}~\bibnamefont{Ghosh}}, \bibinfo{journal}{Acta
  Mat.} \textbf{\bibinfo{volume}{55}}, \bibinfo{pages}{3347}
  (\bibinfo{year}{2007}).

\bibitem[{\citenamefont{Morgan et~al.}(2005)\citenamefont{Morgan, Ceder, and
  Curtarolo}}]{morgan:meas_2005_ht}
\bibinfo{author}{\bibfnamefont{D.}~\bibnamefont{Morgan}},
  \bibinfo{author}{\bibfnamefont{G.}~\bibnamefont{Ceder}}, \bibnamefont{and}
  \bibinfo{author}{\bibfnamefont{S.}~\bibnamefont{Curtarolo}},
  \bibinfo{journal}{Meas. Sci. Technolog.} \textbf{\bibinfo{volume}{16}},
  \bibinfo{pages}{296} (\bibinfo{year}{2005}).

\bibitem[{\citenamefont{Curtarolo et~al.}(2009)\citenamefont{Curtarolo, Hart,
  Setyawan, Chepulskii, Levy, and Morgan}}]{aflow}
\bibinfo{author}{\bibfnamefont{S.}~\bibnamefont{Curtarolo}},
  \bibinfo{author}{\bibfnamefont{G.~L.~W.} \bibnamefont{Hart}},
  \bibinfo{author}{\bibfnamefont{W.}~\bibnamefont{Setyawan}},
  \bibinfo{author}{\bibfnamefont{R.~V.} \bibnamefont{Chepulskii}},
  \bibinfo{author}{\bibfnamefont{O.}~\bibnamefont{Levy}}, \bibnamefont{and}
  \bibinfo{author}{\bibfnamefont{D.}~\bibnamefont{Morgan}},
  \bibinfo{journal}{{\it ``AFLOW: software for high-throughput calculation of
  material properties''}, {\sf http://materials.duke.edu/aflow.html}}
  (\bibinfo{year}{2009}).

\bibitem[{\citenamefont{Kresse and Hafner}(1993)}]{kresse_vasp}
\bibinfo{author}{\bibfnamefont{G.}~\bibnamefont{Kresse}} \bibnamefont{and}
  \bibinfo{author}{\bibfnamefont{J.}~\bibnamefont{Hafner}},
  \bibinfo{journal}{Phys.\ Rev.\ B} \textbf{\bibinfo{volume}{47}},
  \bibinfo{pages}{558} (\bibinfo{year}{1993}).

\bibitem[{\citenamefont{Blochl}(1994)}]{PAW}
\bibinfo{author}{\bibfnamefont{P.~E.} \bibnamefont{Blochl}},
  \bibinfo{journal}{Phys.\ Rev.\ B} \textbf{\bibinfo{volume}{50}},
  \bibinfo{pages}{17953} (\bibinfo{year}{1994}).

\bibitem[{\citenamefont{Perdew et~al.}(1996)\citenamefont{Perdew, Burke, and
  Ernzerhof}}]{PBE}
\bibinfo{author}{\bibfnamefont{J.~P.} \bibnamefont{Perdew}},
  \bibinfo{author}{\bibfnamefont{K.}~\bibnamefont{Burke}}, \bibnamefont{and}
  \bibinfo{author}{\bibfnamefont{M.}~\bibnamefont{Ernzerhof}},
  \bibinfo{journal}{Phys.\ Rev.\ Lett.} \textbf{\bibinfo{volume}{77}},
  \bibinfo{pages}{3865} (\bibinfo{year}{1996}).

\bibitem[{\citenamefont{Massalski et~al.}(1990)\citenamefont{Massalski,
  Okamoto, Subramanian, and Kacprzak}}]{Massalski}
\bibinfo{editor}{\bibfnamefont{T.~B.} \bibnamefont{Massalski}},
  \bibinfo{editor}{\bibfnamefont{H.}~\bibnamefont{Okamoto}},
  \bibinfo{editor}{\bibfnamefont{P.~R.} \bibnamefont{Subramanian}},
  \bibnamefont{and} \bibinfo{editor}{\bibfnamefont{L.}~\bibnamefont{Kacprzak}},
  eds., \emph{\bibinfo{title}{Binary alloy phase diagrams}}
  (\bibinfo{publisher}{American Society for Metals}, \bibinfo{year}{1990}).

\bibitem[{\citenamefont{Villars et~al.}(2004)\citenamefont{Villars, Berndt,
  Brandenburg, Cenzual, Daams, Hulliger, Massalski, Okamoto, Osaki, Prince
  et~al.}}]{Pauling}
\bibinfo{author}{\bibfnamefont{P.}~\bibnamefont{Villars}},
  \bibinfo{author}{\bibfnamefont{M.}~\bibnamefont{Berndt}},
  \bibinfo{author}{\bibfnamefont{K.}~\bibnamefont{Brandenburg}},
  \bibinfo{author}{\bibfnamefont{K.}~\bibnamefont{Cenzual}},
  \bibinfo{author}{\bibfnamefont{J.}~\bibnamefont{Daams}},
  \bibinfo{author}{\bibfnamefont{F.}~\bibnamefont{Hulliger}},
  \bibinfo{author}{\bibfnamefont{T.}~\bibnamefont{Massalski}},
  \bibinfo{author}{\bibfnamefont{H.}~\bibnamefont{Okamoto}},
  \bibinfo{author}{\bibfnamefont{K.}~\bibnamefont{Osaki}},
  \bibinfo{author}{\bibfnamefont{A.}~\bibnamefont{Prince}},
  \bibnamefont{et~al.}, \bibinfo{journal}{Journal of Alloys and Compounds}
  \textbf{\bibinfo{volume}{367}}, \bibinfo{pages}{293} (\bibinfo{year}{2004}).

\bibitem[{\citenamefont{Kolmogorov and
  Curtarolo}(2006{\natexlab{a}})}]{kolmogorov:binary_LiB_2006}
\bibinfo{author}{\bibfnamefont{A.~N.} \bibnamefont{Kolmogorov}}
  \bibnamefont{and}
  \bibinfo{author}{\bibfnamefont{S.}~\bibnamefont{Curtarolo}},
  \bibinfo{journal}{Phys.\ Rev.\ B} \textbf{\bibinfo{volume}{73}},
  \bibinfo{pages}{180501(R)} (\bibinfo{year}{2006}{\natexlab{a}}).

\bibitem[{\citenamefont{Kolmogorov and
  Curtarolo}(2006{\natexlab{b}})}]{kolmogorov:binary_borides_LiB_2006}
\bibinfo{author}{\bibfnamefont{A.~N.} \bibnamefont{Kolmogorov}}
  \bibnamefont{and}
  \bibinfo{author}{\bibfnamefont{S.}~\bibnamefont{Curtarolo}},
  \bibinfo{journal}{Phys.\ Rev.\ B} \textbf{\bibinfo{volume}{74}},
  \bibinfo{pages}{224507} (\bibinfo{year}{2006}{\natexlab{b}}).

\bibitem[{\citenamefont{Hart and Forcade}(2008)}]{gus_enum}
\bibinfo{author}{\bibfnamefont{G.~L.~W.} \bibnamefont{Hart}} \bibnamefont{and}
  \bibinfo{author}{\bibfnamefont{R.~W.} \bibnamefont{Forcade}},
  \bibinfo{journal}{Phys.\ Rev.\ B} \textbf{\bibinfo{volume}{77}},
  \bibinfo{pages}{224115} (\bibinfo{year}{2008}).

\bibitem[{\citenamefont{Pettifor}(1984)}]{pettifor:1984}
\bibinfo{author}{\bibfnamefont{D.~G.} \bibnamefont{Pettifor}},
  \bibinfo{journal}{Sol. State Commun.} \textbf{\bibinfo{volume}{51}},
  \bibinfo{pages}{31} (\bibinfo{year}{1984}).

\bibitem[{\citenamefont{Pettifor}(1986)}]{pettifor:1986}
\bibinfo{author}{\bibfnamefont{D.~G.} \bibnamefont{Pettifor}},
  \bibinfo{journal}{J. Phys. C: Solid State Phys.}
  \textbf{\bibinfo{volume}{19}}, \bibinfo{pages}{285} (\bibinfo{year}{1986}).

\bibitem[{\citenamefont{{{v}an {d}e Walle} and Ceder}(2002)}]{Axel_RMP}
\bibinfo{author}{\bibfnamefont{A.}~\bibnamefont{{{v}an {d}e Walle}}}
  \bibnamefont{and} \bibinfo{author}{\bibfnamefont{G.}~\bibnamefont{Ceder}},
  \bibinfo{journal}{Rev. Mod. Phys.} \textbf{\bibinfo{volume}{74}},
  \bibinfo{pages}{11} (\bibinfo{year}{2002}).

\bibitem[{\citenamefont{Wolverton and
  Ozoli\c{n}\v{s}}(2001)}]{wolverton:prl_2001_AlCu}
\bibinfo{author}{\bibfnamefont{C.}~\bibnamefont{Wolverton}} \bibnamefont{and}
  \bibinfo{author}{\bibfnamefont{V.}~\bibnamefont{Ozoli\c{n}\v{s}}},
  \bibinfo{journal}{Phys.\ Rev.\ Lett.} \textbf{\bibinfo{volume}{86}},
  \bibinfo{pages}{5518} (\bibinfo{year}{2001}).

\end{thebibliography}

\end{document}